\documentclass[11pt,draft]{article}
\usepackage{amssymb,amsmath}
\usepackage[T2A]{fontenc}
\usepackage[]{algorithm2e}
\usepackage[cp1251]{inputenc}
\usepackage{textcomp}
\usepackage[russian]{babel}
\usepackage[dvips]{graphicx}
\usepackage[usenames, dvipsnames]{color}
\newcounter{proposition}

\newcommand{\nothing}[1]{}

\newcommand{\beq}[1]{\begin{equation}\label{#1}}
\newcommand{\eeq}{\end{equation}}
\newcommand{\req}[1]{(\ref{#1})}
\newcommand{\bmu}[1]{\begin{multline}\label{#1}}
\newcommand{\emu}{\end{multline}}

\newcommand{\eq}{\triangleq}

\renewcommand{\varlimsup}{\mathop{\overline{\lim}}\limits}

\newcommand{\x}{{\textbf{\textit{x}}}}
\newcommand{\y}{{\textbf{\textit{y}}}}
\renewcommand{\a}{{\textbf{\textit{a}}}}
\renewcommand{\b}{{\textbf{\textit{b}}}}

\renewcommand{\a}{{\bf a}}
\renewcommand{\b}{{\bf b}}

\renewcommand{\u}{{\bf u}}
\renewcommand{\v}{{\bf v}}
\renewcommand{\S}{{\cal S}}

\renewcommand{\L}{{\bf L}}

\renewcommand{\l}{\ell}

\renewcommand{\S}{{\mathcal{S}}}
\renewcommand{\L}{{\mathcal{L}}}

\newcommand{\s}{{ {s}}}

\renewcommand{\chi}{\upsilon}
\renewcommand{\l}{{ {L}}}
\newcommand{\0}{{\textbf{\textit{0}}}}
\newcommand{\1}{{\textbf{\textit{1}}}}
\renewcommand{\(}{\left(}
\renewcommand{\)}{\right)}
\renewcommand{\[}{\left[}
\renewcommand{\]}{\right]}
\newcommand{\len}{\left\lfloor}
\newcommand{\rin}{\right\rfloor}

\renewcommand{\l}{\ell}
\renewcommand{\No}{\textnumero}

\begin{document}


\begin{center}
   {\bf А.Г. Дьячков, И.В. Воробьев, Н.А. Полянский, В.Ю. Щукин}
\end{center}

\begin{center}
{\Large\bf  Границы скорости  дизъюнктивных кодов}
\end{center}

\begin{abstract}
Данная статья является исправленной версией статьи \cite{d14_pit}, в которой по вине авторов при доказательстве теоремы 4 была допущена существенная ошибка. Статья содержит более слабые, но представляющие для читателя интерес утверждения о границах скорости кодов свободных от перекрытий, которые получаются с помощью метода, применяемого в \cite{d14_pit} при выводе теоремы 4.

Двоичный код называется дизъюнктивным  свободным от перекрытий  $(s,\ell)$-кодом (СП  $(s,\ell)$-кодом),
если  он является матрицей
инцидентности семейства множеств, для которого пересечение любых $\ell$ множеств
не покрывается  объединением $s$ любых других множеств данного семейства.
Двоичный код называется дизъюнктивным кодом со списочным декодированием силы $s$ с объемом списка $L$ (СД  $s_L$-кодом),
если он является матрицей инцидентности семейства множеств,
для которого объединение любых $s$ множеств может покрывать не более $L-1$
других множеств данного семейства.
При $L=\ell=1$ оба определения совпадают, и соответствующий двоичный код
называется дизъюнктивным $s$-кодом. Цель настоящей работы - уточнение ранее известных и
получение новых границ для скоростей данных кодов. В частности, с помощью метода случайного кодирования на ансамбле двоичных равновесных
кодов получены более точные нижние границы скорости СП $(s,\ell)$-кодов и СД $s_L$-кодов. Также в работе установлена верхняя граница скорости  СД $s_L$-кодов.
 \end{abstract}

\section{Обозначения, определения и  результаты}
\quad
Пусть $N$, $t$, $s$, $L$ и $\ell$ - целые числа,
 $1\le s<t$, $1\le L\le t-s$,  $1\le \ell\le t-s$, символ
$\eq$ обозначает равенство по определению, $|A|$ -- объем множества $A$, а
$[N]\eq\{1,2,\dots,N\}$ - множество целых чисел от $1$ до~$N$. Введем
\beq{X}
X=\|x_i(j)\|, \; x_i(j)=0,1,\;\x_i\eq(x_i(1),\dots,x_i(t)),\;
\x(j)\eq(x_1(j),\dots,x_N(j)),
\eeq
 $i\in[N]$, $j\in[t]$, -- двоичную $(N\times t)$-матрицу с $N$ строками $\x_1,\dots,\x_N$ и $t$ столбцами
$\x(1),\dots,\x(t)$ (кодовыми словами), которую далее будем называть
{\em  кодом длины $N$  и объема $t$}. Число единиц в столбце  $x(j)$, т.е.
$|\x(j)|\eq\sum\limits_{i=1}^N\,x_i(j)$, будем называть {\em весом} столбца  $\x(j)$,~$j\in[t]$.
Будем говорить, что код $X$ является {\em равновесным}, если каждое его кодовое слово содержит
одинаковое число $w$, $1\le w<N$, единиц,
т.е. вес~$|\x(j)|=w$ для любого~$j\in[t]$.
Стандартный символ  $\bigvee$ обозначает
операцию дизъюнктивной (булевой) суммы двух двоичных чисел
$$
0\bigvee0=0,\quad  0\bigvee 1=1\bigvee 0 = 1\bigvee 1 = 1,
$$
а также покомпонентную  дизъюнктивную  сумму двух  двоичных столбцов.
Будем говорить, что столбец $\u$ покрывает столбец $\v$ ($\u \succeq \v$), если
$\u\bigvee \v=\u$. Стандартный символ $\lfloor a\rfloor$
($\lceil a\rceil$) используем для обозначения  наибольшего (наименьшего) целого числа $\le a$~($\ge a$).

\textbf{Определение 1.}~\cite{mp88}.\quad
Код $X$ называется {\em дизъюнктивным свободным от перекрытий $(s,\ell)$-кодом}
(кратко, {\em СП $(s,\ell)$-кодом}), если для любых двух
непересекающихся  множеств $\S,\,\L\subset[t]$, $|\S|=\s$, $|\L|=\ell$, $\S\cap\L=\varnothing$,
существует строка $\x_i$, $i\in [N]$, для которой
$$
 x_i(j)=0 \quad \text{для любого}\,j\in\S,\quad \text{а}\quad  x_i(k)=1\quad \text{для любого}\quad k\in\L.
$$
Учитывая очевидную симметрию по $s$  и $\l$, обозначим через
$t_{cf}(N,s,\ell)=t_{cf}(N,\ell,s)$ - максимальный объем
СП $(s,\ell)$-кодов длины $N$, а через
$N_{cf}(t,s,\ell)=N_{cf}(t,\ell,s)$ обозначим
минимальное число строк СП $(s,\ell)$-кодов объема $t$  и
определим {\em скорость} СП $(s,\ell)$-кодов:
 \beq{Rsl}
 R(s,\ell)=R(\ell,s)\eq \varlimsup_{N\to\infty}\frac{\log_2 t_{cf}(N,s,\ell)}{N}\,=\,
 \varlimsup_{t\to\infty}\frac{\log_2 t}{N_{cf}(t,\s,\ell)}.
 \eeq

\textbf{Определение 2.}~\cite{dr81},\cite{dr83}.\quad
Код $X$ называется {\em дизъюнктивным кодом со списочным декодированием}
 {\em силы $s$ с объемом списка $L$} (кратко, {\em СД  $\;s_L$-кодом}), если дизъюнктивная сумма
любых $s$ столбцов кода $X$ покрывает не более $L-1$ других
столбцов кода $X$, не входящих  в эту сумму.
Обозначим через $t_{ld}(N,s,L)$ -- максимальный объем
СД $\;s_L$-кодов длины $N$, а через $N_{ld}(t,s,L)$ -- минимальное число строк СД $\;s_L$-кодов объема $t$ и
определим {\em скорость} СД $\;s_L$-кодов:
 \beq{RsL}
 R_L(s)\eq \varlimsup_{N\to\infty}\frac{\log_2 t_{ld}(N,s,L)}{N}\,=\,
  \varlimsup_{t\to\infty}\frac{\log_2 t}{N_{ld}(t,s,L)}.
 \eeq

При $L=\ell=1$ определения 1 и 2 совпадают,  скорость $R_1(s)=R(s,1)$, $s=1,2,\dots$, и соответствующий
код называется   {\em дизъюнктивным $s$-кодом}. Дизъюнктивные $s$-коды   были введены в 1964 году в основополагающей
статье Каутса и Синглтона~\cite{ks64}, где
были также установлены  первые нетривиальные свойства дизъюнктивных $s$-кодов, разработаны их важные приложения и конструкции,
которые в дальнейшем существенно развивались
в~\cite{dmr00_1}-\cite{dmr00_2},
а также была поставлена задача получения границ скорости~$R(s,1)$.

\subsection{Нижняя и верхняя границы скорости $R(s,1)$}
\quad
Наилучшая к настоящему времени нижняя граница скорости
$R(s,1)$ была получена в 1989 году в работе~\cite{drr89},
в которой с помощью метода случайного кодирования на ансамбле двоичных равновесных
кодов\footnote[1]{Ансамбль равновесных кодов представляет собой
частный случай ансамбля кодов, называемого ансамблем с фиксированной
композицией (АФК), который был введен в  книге~\cite{f65} при построении методом случайного кодирования
наилучших верхних границ  вероятности ошибки для дискретных каналов без памяти.
В книге~\cite{ck85} АФК применялся также в аналогичной задаче для
каналов  без памяти с множественным доступом.
В работах~\cite{d80}-\cite{dr89} исследовалась логарифмическая
асимптотика средней по АФК вероятности ошибки.}
показано, что
\beq{ran1}
R(s,1)\ge\underline{R}(s,1)\eq\,s^{-1}\cdot\,\max\limits_{0<Q<1}\,{A(s,Q)},\qquad s=1,2,\dots,
\eeq
\beq{ran2}
 A(s,Q)\eq\log_2\frac{Q}{1-y}-sK(Q,1-y)-\,K\left(Q,\frac{1-y}{1-y^s}\right),
\eeq
где используется стандартное обозначение расстояния Кульбака
\beq{Kul}
K(a,b)\eq a\cdot \log_2 \frac{a}{b}+(1-a)\cdot \log_2 \frac{1-a}{1-b},\quad 0<a,b<1,
\eeq
 а  $y=y(s,Q)$, $1-Q\le y<1$, -- единственный корень уравнения
\beq{ran3}
y=1\,-\,Q\,+\,Q\,y^s\cdot\frac{1-y}{1-y^s},
\qquad 1-Q\le y<1.
\eeq
Отметим, что  граница случайного кодирования $\underline{R}(s,1)$, установленная  в~\cite{drr89},
эквивалентна~\req{ran1}-\req{ran3}, но  отличается по форме записи.

Если $s\to\infty$, то асимптотика границы~\req{ran1}-\req{ran3} имеет вид
\beq{ran4}
R(s,1)\,\ge\,\underline{R}(s,1)=\frac{1}{s^2\log_2 e }(1+o(1))=
\frac{0,693}{s^2}(1+o(1)).
\eeq
где здесь и далее $e=2,718$ -- основание натурального логарифма.

Очевидно~\cite{ks64}, что  $R(s,1)\le 1/s$, $s=1,2,\dots$, а нетривиальная
верхняя граница скорости $R(s,1)$, которая до настоящего времени является
наилучшей,  была построена в 1982 году в работе~\cite{dr82}.
Для описания этой границы,  обозначаемой в данной работе символом
$\overline{R}(s,1)$, $s=1,2,\dots$, и называемой {\em рекуррентной границей}, введем
стандартное обозначение двоичной энтропии
\beq{entropy}
h(v)\,\eq-v\,\log_2v-(1-v)\log_2(1-v), \qquad 0<v<1,
\eeq
и функцию
\beq{rec1}
f_s(v)\eq h(v/s)-v\, h(1/s),\quad 0<v<1,\quad s=1,2,\dots,
\eeq
аргумента $v$, $0<v<1$. В~\cite{dr82} показано (см. также~\cite{dmtv02}), что
функция $f_s(v)>0$,  выпукла вверх и принимает максимальное значение:
\beq{rec1-1}
\max\limits_{0<v<1}\;f_s(v)\,=\,f_s(v_s)\quad\text{при}\quad
v_s\eq\frac{s}{1+2^{s\cdot h\left(\frac{1}{s}\right)}},\quad s=1,2,\dots.
\eeq
Положим
\beq{rec2}
\overline{R}(1,1)\eq 1,\quad
\overline{R}(2,1)\eq\max_{0<v<1}\;f_2(v)\,=\,f_2(v_2)\,=\,0,322,
\eeq
а далее последовательность $\overline{R}(s,1)$, $s=3,4,\dots$, определяется~\cite{dr82} как
единственное решение рекуррентного уравнения
\beq{rec3}
\overline{R}(s,1)=\,f_s\left(1-
\frac{\overline{R}(s,1)}{\overline{R}(s-1,1)}\right),\quad s=3,4,\dots.
\eeq

Для скорости $R(s,1)$ и
рекуррентной последовательности  $\overline{R}(s,1)$, $s=1,2,\dots$,
описываемой~\req{rec2}-\req{rec3}, в~\cite{dr82} были доказаны неравенства
\beq{up-non-rec}
R(s,1)\,\le\,\overline{R}(s,1)\,\le\,\frac{2\log_2[e(s+1)/2]}{s^2},\qquad s=2,3,\dots,
\eeq
которые давали асимптотическую верхнюю границу
для скорости $R(s,1)$:
\beq{upR1-as}
R(s,1)\,\le\,\frac{2\log_2s}{s^2}\,(1+o(1)), \quad
s\to\infty.
\eeq
При $s=2,3,\dots,6$ численные значения нижней  границы $\underline{R}(s,1)$,
из~\req{ran1}-\req{ran3}, и  верхней границы $\overline{R}(s,1)$
из~\req{rec2}-\req{rec3}, внесены в сводную Таблицу~1.

В разделе 2.1  будет доказана

\textbf{Теорема 1.} \quad
{\em Если $s\ge8$, то   рекуррентная последовательности  $\overline{R}(s,1)$,
определяемая~$\req{rec2}-\req{rec3}$, удовлетворяет
неравенству}
\beq{low-non-rec}
\overline{R}(s,1)\,\ge\,\frac{2\log_2[(s+1)/8]}{(s+1)^2},\qquad s\ge8.
\eeq

Из~\req{up-non-rec} и~\req{low-non-rec} вытекает асимптотическое равенство
\beq{rec-as}
\overline{R}(s,1)\,=\,\frac{2\log_2s}{s^2}\,(1+o(1)), \quad
s\to\infty.
\eeq

\subsection{Границы  скорости $ R(s,\ell)$ для СП $(s,\ell)$-кодов, $2\le\ell\le s$ }
\quad
Дизъюнктивные свободные от перекрытий $(s,\ell)$-коды (СП $(s,\ell)$-коды)
были  введены  в 1988 году в статье~\cite{mp88} в связи с криптографической
задачей распределения ключей, описание которой можно также  найти в статьях~\cite{l02}-\cite{l03}.
Биологические приложения и конструкции  СП $(s,\ell)$-кодов в задачах неадаптивного
 группового тестирования при поиске супермножеств (комплексов)  предложены в работах~\cite{dmtv02}
и~\cite{dmtvy00}-\cite{dmtv01} .
Значимое приложение СП $(s-\ell+1,\ell)$-кодов, $2\le \ell<s$, для $\ell$-пороговой модели
неадаптивных групповых проверок~\cite{d13} при поиске $\le s$  дефектов отмечено в~\cite{cf09}.
Последующие конструкции СП $(s,\ell)$-кодов были разработаны в~\cite{kl04}-\cite{sp09}.

Первые результаты исследования верхних границ скорости $R(s,\ell)$ для СП $(s,\ell)$-кодов,
$2\le\ell\le s$, были получены в~\cite{dmtv02} и~\cite{s00}.
В работах~\cite{l02}-\cite{l03} для $R(s,\l)$ было доказано  неравенство
\beq{leb}
R(s,\l)\,\le\;
\frac{R(s-i,\l-j)}{R(s-i,\l-j)+\frac{(i+j)^{i+j}}{i^i\cdot j^j}},\quad i\in[s-1],\; j\in[\l-1],
\eeq
которое представляет собой уточнение неравенства
\beq{eng}
R(s,\l)\,\le\, R(s-i,\l-j)\cdot\frac{i^i\cdot j^j}{(i+j)^{i+j}},\quad i\in[s-1],\; j\in[\l-1],
\eeq
ранее установленного в~\cite{e96}.
Рекуррентное неравенство~\req{leb}  и рекуррентная верхняя граница
$\overline{R}(s,1)$, $s\ge1$, определяемая~\req{rec1}-\req{rec3}, дают
для скорости~$R(s,\l)$, $2\le\l\le s$ наилучшую известную верхнюю рекуррентную границу:
\beq{rec-sl}
R(s,\l)\le\overline{R}(s,\l)\eq\min\limits_{i\in[s-1]}\min\limits_{j\in[\l-1]}\,
\frac{\overline{R}(s-i,\l-j)}{\overline{R}(s-i,\l-j)+\frac{(i+j)^{i+j}}{i^i\cdot j^j}}.
\eeq
Асимптотику  границы~\req{rec-sl}
описывает представленная ранее в докладе~\cite{dvy02}
и доказываемая  в разделе~2.2

\textbf{Теорема 3.}~\cite{dvy02}. \quad {\em Если $s\to\infty$ и $\l\ge2$ фиксировано, то}
\beq{rec-sl-as}
R(s,\l)\,\le\,\overline{R}(s,\l)\,\le\,
\frac{(\l+1)^{\l+1}}{2\,e^{\l-1}}\cdot
\frac{\log_2\,s}{s^{\l+1}}\,\cdot(1+o(1)),\quad \l\ge2,\quad s\to\infty.
\eeq
\medskip

Наилучшая в настоящее время нижняя граница скорости
$R(s,\l)$, $2\le\l\le s$, была получена в 2002 году в статье~\cite{dmtv02} с
помощью метода случайного кодирования на  ансамбле с независимыми
двоичными компонентами кодовых слов и предложенном в~\cite{nz88} некотором специальном
 ансамбле с независимыми двоичными равновесными словами.
При фиксированном $\l\ge2$ и  $s\to\infty$  асимптотика этой границы имела вид
\beq{Low02-as}
R(s,\l)\,\ge\,\frac{e^{-\ell}\ell^{\ell}\log_2{e}}{s^{\ell+1}}(1+o(1)),\quad \l\ge2,\quad s\to\infty.
\eeq

 Одним из важных результатов данной статьи является доказываемая в разделе~2.2
теорема~4, в которой с помощью метода случайного кодирования на ансамбле равновесных
двоичных кодов найдена нижняя граница $\underline{R}(s,\ell)$ для скорости $R(s,\l)$, $2\le\l\le s$,
и исследована асимптотика функции $\underline{R}(s,\ell)$, когда $s\to\infty$ и $\l\ge2$ фиксировано.

\textbf{Теорема 4.}\quad (Граница случайного кодирования $\underline{R}(s,\ell)$.)\textit{
Имеют место следующие два утверждения.\\
{\bf1.} \quad
Пусть $2\le\l\le s$. Тогда  скорость СП $(\s,\ell)$-кодов
\beq {Th4}
R(s,\ell)\ge \underline{R}(s,\ell)\,\eq\,\frac{1}{s+\ell-1}\max\limits_{\substack{\req{LINKzu}\\0<z,\, u<1}}\,T(z,u,s,\ell),               
\eeq
где функция $T(z,u,s,\ell)$ определена следующим образом
\begin{multline}\label{T}
T(z,u,s,\ell)\eq \frac{su}{1-(z-u)}\log_2\[\frac{z}{u}\] + \frac{\l(1-z)}{1- (z-u)}\log_2\[\frac{1-u}{1-z}\] +\\
+ (s+\l -1 )\log_2\[1- (z-u)\],
\end{multline}
а параметры $z$ и $u$, $0<z,\,u<1$, связаны между собой посредством следующего равенства
\beq{LINKzu}
z-u = z^s(1-u)^\l.
\eeq
{\bf2.} \quad
Если $s\to\infty$ и $\l\ge2$ фиксировано, то
для нижней границы $\underline{R}(s,\ell)$ справедливо асимптотическое равенство
\beq{Con1}
{R}(s,\ell)\ge\underline{R}(s,\ell)=\frac{e^{-\ell}\ell^{\ell}\log_2{e}}{s^{\ell+1}}(1+o(1)),
\quad  \l=2,3,\dots, \quad  s\to\infty.
\eeq
}
 При доказательстве
 утверждения~2 теоремы~4 для оптимального веса $Q(s,\ell)$ кодовых слов для ансамбля равновесных двоичных кодов будет установлено асимптотическое равенство
 \beq{Qls}
 Q(s,\ell)=\frac{\l}{s}\,(1+o(1)),\quad s\to\infty,\quad \l=2,3,\dots.
 \eeq
 
\textbf{Замечание 1.}\quad Отметим, что численные значения, полученные с помощью Теоремы 4 и отраженные в Таблице 1, улучшают ранее известные нижние границы для $R(s,\ell)$ (ср. с Таблицей 2 в \cite{dmtv02}); в то же время асимптотическое поведение $\underline{R}(s,\ell)$, установленное в Теореме 4, в точности совпадает с ранее полученным результатом в \cite{dmtv02}.
 
В Таблице~1 при $\l=1$ и $2\le s\le 10$ даны числовые значения верхней границы \req{rec3}, а также нижней границы \req{ran1}  для $R(s,1)$  вместе с долeй $Q(s)$ оптимального веса кодовых слов для ансамбля равновесных двоичных кодов. 
 При   $1\le\l\le s\le 10$, в  сводной  Таблице 1 также
 указаны  верхняя граница  $\overline{R}(s,\l)$,
 определяемая правой частью~\req{rec-sl},
 нижняя граница $\underline{R}(s,\ell)$ и соответствующая
 доля $Q(s,\ell)$ оптимального веса кодовых слов для ансамбля равновесных двоичных кодов
 в границе случайного кодирования из теоремы 4.
 \begin{table}[h!]\label{TableRsl}
 	\caption{}\label{tab2}
 	\begin{tabular*}{\textwidth}{@{\extracolsep{\fill}}ccccccc}
 		\hline
 		$(s,\ell)$ & $(2,1)$  & $(3,1)$  & $(4,1)$  & $(5,1)$  & $(6,1)$  & $(7,1)$ \cr 
 		$\overline{R}(s,\ell)$  & $3.22 \cdot 10^{-1}$ & $1.99 \cdot 10^{-1}$ & $1.40 \cdot 10^{-1}$ & $1.06 \cdot 10^{-1}$ & $8.30 \cdot 10^{-2}$ & $6.73 \cdot 10^{-2}$\cr 
 		$\underline{R}(s,\ell)$  & $1.83 \cdot 10^{-1}$ & $7.87 \cdot 10^{-2}$ & $4.39 \cdot 10^{-2}$ & $2.79 \cdot 10^{-2}$ & $1.94 \cdot 10^{-2}$ & $1.42 \cdot 10^{-2}$\cr 
 		$Q(s,\ell)$ & $0.26$ & $0.19$ & $0.15$ & $0.12$ & $0.10$ & $0.09$\cr 
 		\hline
 		$(s,\ell)$ & $(8,1)$  & $(9,1)$  & $(10,1)$  & $(2,2)$  & $(3,2)$  & $(4,2)$ \cr 
 		$\overline{R}(s,\ell)$  & $5.59 \cdot 10^{-2}$ & $4.73 \cdot 10^{-2}$ & $4.07 \cdot 10^{-2}$ & $1.61 \cdot 10^{-1}$ & $7.45 \cdot 10^{-2}$ & $4.55 \cdot 10^{-2}$\cr 
 		$\underline{R}(s,\ell)$  & $1.09 \cdot 10^{-2}$ & $8.58 \cdot 10^{-3}$ & $6.94 \cdot 10^{-3}$ & $3.66 \cdot 10^{-2}$ & $1.41 \cdot 10^{-2}$ & $6.90 \cdot 10^{-3}$\cr 
 		$Q(s,\ell)$ & $0.08$ & $0.07$ & $0.06$ & $0.50$ & $0.40$ & $0.33$\cr 
 		\hline
 		$(s,\ell)$ & $(5,2)$  & $(6,2)$  & $(7,2)$  & $(8,2)$  & $(9,2)$  & $(10,2)$ \cr 
 		$\overline{R}(s,\ell)$  & $2.87 \cdot 10^{-2}$ & $2.04 \cdot 10^{-2}$ & $1.46 \cdot 10^{-2}$ & $1.10 \cdot 10^{-2}$ & $8.58 \cdot 10^{-3}$ & $6.75 \cdot 10^{-3}$\cr 
 		$\underline{R}(s,\ell)$  & $3.90 \cdot 10^{-3}$ & $2.42 \cdot 10^{-3}$ & $1.60 \cdot 10^{-3}$ & $1.12 \cdot 10^{-3}$ & $8.11 \cdot 10^{-4}$ & $6.06 \cdot 10^{-4}$\cr 
 		$Q(s,\ell)$ & $0.28$ & $0.24$ & $0.22$ & $0.20$ & $0.18$ & $0.16$\cr 
 		\hline
 		$(s,\ell)$ & $(3,3)$  & $(4,3)$  & $(5,3)$  & $(6,3)$  & $(7,3)$  & $(8,3)$ \cr 
 		$\overline{R}(s,\ell)$  & $3.72 \cdot 10^{-2}$ & $1.83 \cdot 10^{-2}$ & $1.09 \cdot 10^{-2}$ & $6.70 \cdot 10^{-3}$ & $4.23 \cdot 10^{-3}$ & $3.01 \cdot 10^{-3}$\cr 
 		$\underline{R}(s,\ell)$  & $4.78 \cdot 10^{-3}$ & $2.09 \cdot 10^{-3}$ & $1.06 \cdot 10^{-3}$ & $5.96 \cdot 10^{-4}$ & $3.61 \cdot 10^{-4}$ & $2.31 \cdot 10^{-4}$\cr 
 		$Q(s,\ell)$ & $0.50$ & $0.42$ & $0.37$ & $0.33$ & $0.30$ & $0.27$\cr 
 		\hline
 		$(s,\ell)$ & $(9,3)$  & $(10,3)$  & $(4,4)$  & $(5,4)$  & $(6,4)$  & $(7,4)$ \cr 
 		$\overline{R}(s,\ell)$  & $2.13 \cdot 10^{-3}$ & $1.54 \cdot 10^{-3}$ & $9.14 \cdot 10^{-3}$ & $4.55 \cdot 10^{-3}$ & $2.57 \cdot 10^{-3}$ & $1.57 \cdot 10^{-3}$\cr 
 		$\underline{R}(s,\ell)$  & $1.55 \cdot 10^{-4}$ & $1.08 \cdot 10^{-4}$ & $8.20 \cdot 10^{-4}$ & $3.76 \cdot 10^{-4}$ & $1.93 \cdot 10^{-4}$ & $1.07 \cdot 10^{-4}$\cr 
 		$Q(s,\ell)$ & $0.25$ & $0.23$ & $0.50$ & $0.44$ & $0.40$ & $0.36$\cr 
 		\hline
 		$(s,\ell)$ & $(8,4)$  & $(9,4)$  & $(10,4)$  & $(5,5)$  & $(6,5)$  & $(7,5)$ \cr 
 		$\overline{R}(s,\ell)$  & $9.90 \cdot 10^{-4}$ & $6.26 \cdot 10^{-4}$ & $4.35 \cdot 10^{-4}$ & $2.27 \cdot 10^{-3}$ & $1.14 \cdot 10^{-3}$ & $6.25 \cdot 10^{-4}$\cr 
 		$\underline{R}(s,\ell)$  & $6.34 \cdot 10^{-5}$ & $3.95 \cdot 10^{-5}$ & $2.56 \cdot 10^{-5}$ & $1.57 \cdot 10^{-4}$ & $7.39 \cdot 10^{-5}$ & $3.79 \cdot 10^{-5}$\cr 
 		$Q(s,\ell)$ & $0.33$ & $0.31$ & $0.29$ & $0.50$ & $0.45$ & $0.42$\cr 
 		\hline
 		$(s,\ell)$ & $(8,5)$  & $(9,5)$  & $(10,5)$  & $(6,6)$  & $(7,6)$  & $(8,6)$ \cr 
 		$\overline{R}(s,\ell)$  & $3.74 \cdot 10^{-4}$ & $2.29 \cdot 10^{-4}$ & $1.44 \cdot 10^{-4}$ & $5.68 \cdot 10^{-4}$ & $2.84 \cdot 10^{-4}$ & $1.53 \cdot 10^{-4}$\cr 
 		$\underline{R}(s,\ell)$  & $2.08 \cdot 10^{-5}$ & $1.21 \cdot 10^{-5}$ & $7.36 \cdot 10^{-6}$ & $3.21 \cdot 10^{-5}$ & $1.53 \cdot 10^{-5}$ & $7.82 \cdot 10^{-6}$\cr 
 		$Q(s,\ell)$ & $0.38$ & $0.36$ & $0.33$ & $0.50$ & $0.46$ & $0.43$\cr 
 		\hline
 		$(s,\ell)$ & $(9,6)$  & $(10,6)$  & $(7,7)$  & $(8,7)$  & $(9,7)$  & $(10,7)$ \cr 
 		$\overline{R}(s,\ell)$  & $8.87 \cdot 10^{-5}$ & $5.43 \cdot 10^{-5}$ & $1.42 \cdot 10^{-4}$ & $7.10 \cdot 10^{-5}$ & $3.77 \cdot 10^{-5}$ & $2.15 \cdot 10^{-5}$\cr 
 		$\underline{R}(s,\ell)$  & $4.26 \cdot 10^{-6}$ & $2.43 \cdot 10^{-6}$ & $6.78 \cdot 10^{-6}$ & $3.25 \cdot 10^{-6}$ & $1.66 \cdot 10^{-6}$ & $8.98 \cdot 10^{-7}$\cr 
 		$Q(s,\ell)$ & $0.40$ & $0.37$ & $0.50$ & $0.47$ & $0.44$ & $0.41$\cr 
 		\hline
 		$(s,\ell)$ & $(8,8)$  & $(9,8)$  & $(10,8)$  & $(9,9)$  & $(10,9)$  & $(10,10)$ \cr 
 		$\overline{R}(s,\ell)$  & $3.55 \cdot 10^{-5}$ & $1.77 \cdot 10^{-5}$ & $9.34 \cdot 10^{-6}$ & $8.87 \cdot 10^{-6}$ & $4.44 \cdot 10^{-6}$ & $2.22 \cdot 10^{-6}$\cr 
 		$\underline{R}(s,\ell)$  & $1.47 \cdot 10^{-6}$ & $7.09 \cdot 10^{-7}$ & $3.62 \cdot 10^{-7}$ & $3.24 \cdot 10^{-7}$ & $1.57 \cdot 10^{-7}$ & $7.24 \cdot 10^{-8}$\cr 
 		$Q(s,\ell)$ & $0.50$ & $0.47$ & $0.44$ & $0.50$ & $0.47$ & $0.50$\cr 
 		\hline
 	\end{tabular*}
 \end{table}

\subsection{Границы  скорости $R_L(s)$ для СД $\;s_L$-кодов}
\quad
Дизъюнктивные коды со списочным декодированием (СД $\;s_L$-коды)
были  введены в 1981 году в докладе~\cite{dr81} при разработке
системы связи АЛОХА с центральной станцией, когда для различения
сигналов на выходе канала со случайным синхронным множественным доступом
используется кодирование.
Затем некоторые конструкции данных кодов рассматривались в работе~\cite{v98}
(см. также~\cite{dmr00_2} и~\cite{dmtvy00})
в связи с возникающей в молекулярной биологии задачей построения {\em двухступенчатых процедур
групповых проверок} для анализа библиотеки ДНК-клонов. На первой ступени выделяется
множество из  $\le s+L-1$ элементов, которые далее  на второй ступени проверяются поодиночке.
  Отметим, что при фиксированном $s\ge2$ скорость двухступенчатых процедур
  $R_L(s)$ является монотонно неубывающей функцией параметра $L\ge1$ и ее предел
\beq{limR_L}
R_{\infty}(s)\,\eq\,
\lim\limits_{L\to\infty}\,R_L(s)
\eeq
можно интерпретировать как {\em максимальную скорость} двухступенчатых процедур групповых проверок в дизъюнктивной
модели поиска $\le s$  дефектов.

Сформулируем в виде предложений~1-3 важные свойства СД $\;s_L$-кода длины $N$ и объема~$t$,
которые  непосредственно вытекают из его определения~2 с помощью {\em рассуждений от
противного}.
\medskip

\textbf{Предложение 1.}~\cite{dr83}. \quad Пусть $X$ -- произвольный
СД $\;s_L$-код  длины $N$ и объема~$t$.
 Символом $M_s(\y,X)$, $\y\in\{0,1\}^N$,  обозначим множество
всех  $s$-наборов столбцов кода $X$, такое, что для каждого $s$-набора из $M_s(\y,X)$
дизъюнктивная сумма составляющих его $s$ столбцов равна~$\y$.
Рассуждая от противного,  получаем, что для любого $\y\in\{0,1\}^N$ объем
$|M_s(\y,X)|\le {s+L-1\choose s}$. Поскольку число всех  $s$-наборов столбцов кода $X$
равно ${t\choose s}$, то
$$
{t\choose s}\,=
\,\sum_{\y}\left|M_s(\y,X)\right|\,
\le\,{s+L-1\choose s}\,2^N.
$$
Согласно определению~\req{RsL}, для минимального числа строк $N_{ld}(t,s,L)$
(скорости $R_L(s)$) эти неравенства приводят к нижней (верхней)
границе:
\beq{lb-triv1}
N_{ld}(t,s,L)\ge\log_2\frac{{t\choose s}}{{s+L-1\choose s}} \quad\Longrightarrow\quad
 \left(R_L(s)\le \frac{1}{s}\right),\quad s\ge1, \quad L\ge1.
\eeq

\textbf{Предложение 2.}\quad
Пусть $s>L\ge2$, а $X$ есть произвольный СД $\;s_L$-код  длины $N$ и объема~$t$.
Столбец (кодовое слово) кода $X$  назовем  \emph{плохим} для кода $X$, если в $X$ существуют
$\len s/L\rin$ других столбцов, дизъюнктивная сумма которых его покрывает.
В противном случае столбец назовем  \emph{хорошим} и отметим, что множество хороших столбцов
кода $X$ является дизъюнктивным  $\len s/L\rin$-кодом.
Заметим, что  СД $\;s_L$-код $X$ может содержать не более $L-1$ плохих столбцов.
Действительно, если существует $L$-набор плохих столбцов, то он покрывается  дизъюнктивной суммой
не более $L\cdot\len s/L\rin\le s$ столбцов кода, что противоречит определению  СД $\;s_L$-кода.
Другими словами, любой СД $\;s_L$-код  объема~$t$ длины $N$ содержит не менее~$t-(L-1)$ кодовых слов, образующих
 дизъюнктивный  $\len s/L\rin$-код длины $N$. Поэтому для максимального объема
   $t_{ld}(N,s,L)$ и скорости $R_L(s)$ справедливы верхние границы
\beq{lb-triv2}
t_{ld}(N,s,L)\,\le\,t_{ld}\left(N,\len s/L\rin,1\right)+L-1 \quad\Longrightarrow\quad
R_L(s)\,\le\,R\left(\len s/L\rin,1\right),
\quad L\le s.
\eeq

Свойства~\req{lb-triv1}-\req{lb-triv2}
 будут существенно использоваться  в доказательстве верхней границы  теоремы~6
 для скорости $\;R_L(s)$,~$s>L\ge2$.
Эта граница будет построена как обобщение нашей рекуррентной  верхней
границы~\req{rec2}-\req{rec3} для скорости  дизъюнктивных $s$-кодов.
\medskip

\textbf{Предложение 3.}~\cite{dr83}. \quad
{\em Если дизъюнктные  суммы всех $s$-наборов столбцов кода~$X$  отличны друг от друга,
то код $X$ является СД $\;(s-1)_2$-кодом}.\quad
Данное достаточное  условие для СД $\;(s-1)_2$-кода доказывается от противного:
если в коде~$X$ существует набор столбцов, который содержит  $s-1$ столбцов, дизъюнктивная
сумма которых покрывает 2 посторонних столбца, то объединяя данный набор с каждым из этих 2
столбцов, получим два набора, содержащие по $s$ столбцов каждый, дизъюнктивные суммы
которых совпадают.
\medskip

Первые результаты о верхних и нижних границах скорости
$R_L(s)$ при $L\ge2$ были опубликованы в 1983 году в работе~\cite{dr83}.
Верхняя граница получалась как следствие второго неравенства в~\req{lb-triv2}, а
затем для оценки его правой части применялась верхняя граница~\req{up-non-rec}.
Нижняя граница  скорости $R_L(s)$ была построена методом случайного кодирования
для ансамбля кодов с независимыми одинаково распределенными двоичными
компонентами кодовых слов.

В последующих работах~\cite{dra89}-\cite{d03} данные границы уточнялись
и наилучшие к настоящему времени верхняя  и нижняя  границы скорости
$R_L(s)$  сформулированы ниже в виде теоремы~6 и теоремы~7. Эти теоремы были ранее
анонсированы без доказательства в докладе~\cite{dra89}  и обзоре~\cite{d03} и будут доказаны
в разделе~2.3 данной статьи.
\medskip

\textbf{Теорема 6.} \quad (Верхняя рекуррентная граница  $\overline{R}_L(s)$).
{\em Имеют место следующие три утверждения.}\quad
{\bf1.} \quad {\em  Для любого фиксированного $L\ge1$ скорость СД $\;s_L$-кодов удовлетворяет неравенству
$R_L(s)\le\overline{R}_L(s)$, $s=1,2,\dots$,
в правой части  которого последовательность $\overline{R}_L(s)$, $s=1,2,\dots$,
определяется рекуррентно}:
\begin{itemize}
\item
{\em если  $1\le s\le L$, то}
\beq{recL1}
\overline{R}_L(s)\,\eq\,1/s,\qquad  s=1,2,\dots, L\,;
\eeq
\item
{\em если $s\ge L+1$, то
\beq{recL2}
\overline{R}_L(s)\eq\min\{1/s;\,r_L(s)\}, \qquad s=L+1,L+2,\dots,
\eeq
где $r_L(s)$ является единственным решением уравнения
\beq{recL3}
r_L(s)\,\eq\,\max\limits_{\req{recL4}}\,f_{\lfloor s/L\rfloor}(v),\qquad s=L+1,L+2,\dots,
\eeq
в котором  при $n=1,2,\dots$  функция $f_n(v)$ параметра $v$, $0<v<1$, определена~$\req{entropy}-\req{rec1}$ и
максимум берется по всем $v$, удовлетворяющим условию}
\beq{recL4}
0<v<1-\frac{r_L(s)}{\overline{R}_L(s-1)};
\eeq
\item
{\em если   $s>2L$ то уравнение  $\req{recL3}$ можно записать в виде равенства}
\beq{rLs}
r_L(s)=f_{\len s/L\rin}\left(1-\frac{r_L(s)}{\overline{R}_L(s-1)}\right), \quad L\ge1,\quad s>2L.
\eeq
\end{itemize}

{\bf2.}\quad
{\em Для любого $L\ge1$ существует целое число  $s(L)\ge2$, такое, что}
$$
\overline{R}_L(s)=\left\{ \begin{array}{cc}
1/s, & \mbox{если}\;  s=s(L)-1,\\
<1/s, & \mbox{если}\;  s\ge s(L),\\
\end{array}
\right.
$$
{\em и $s(L)=L\log_2 L(1+o(1))$ при} $L~\to\infty$. 

{\bf3.}\quad {\em Если $L\ge 1$ фиксировано и $s\to\infty$, то}
\beq{RLas} \overline{R}_L(s)=\frac{2L\log_2s}{s^{2}} (1+o(1)). \eeq

Определение рекуррентной границы~\req{recL1}-\req{rLs}
и асимптотика~\req{RLas}
представляют собой обобщения рекуррентной границы~\req{rec2}-\req{rec3}
и асимптотики~\req{rec-as}.
\medskip


\textbf{Теорема 7.} \quad (Граница случайного кодирования~$\underline{R}_L(s)$).
{\em Имеют место следующие три утверждения.}\quad
{\bf1.}\quad  {\em Для скорости СД $\;s_L$-кодов
справедливо неравенство}
\beq{ranL1}
R_L(s)\ge\underline{R}_L(s)\,\eq\,\frac{1}{s+L-1}\max\limits_{0<Q<1}\,A_L(s,Q),
\eeq
\beq{ranL2}
 A_L(s,Q)\eq\log_2\frac{Q}{1-y}-sK(Q,1-y)-L\,K\left(Q,\frac{1-y}{1-y^s}\right),
 \quad s\ge1,\; L\ge1,
\eeq
{\em где используется обозначение расстояния Кульбака~$\req{Kul}$, а
параметр $y$, $1-Q\le y<1$, определяется как единственный корень уравнения}
\beq{ranL3}
y=1-Q+Qy^s\left[1-\left(\frac{y-y^s}{1-y^s}\right)^L\right],
\qquad 1-Q\le y<1.
\eeq
{\bf2.}\quad {\em При фиксированном $L=1,2,\dots$ и $s\to\infty$ асимптотика
 границы случайного кодирования имеет вид}
\beq{ranL-4}
\underline{R}_L(s)=\frac{L}{s^2\log_2 e}(1+o(1)).
\eeq
{\bf3.}\quad
{\em При фиксированном $s=2,3,\dots$ и $L\to\infty$ существует предел
\beq{ranL-5}
\underline{R}_{\infty}(s)\,\eq\,\lim\limits_{L\to\infty}\,\underline{R}_{L}(s)=\log_2 \[ \frac{(s-1)^{s-1}}{s^s} + 1 \].
\eeq
Если $s\to\infty$, то данный предел}
$\underline{R}_{\infty}(s)=\frac{\log_2 e}{e\cdot s}(1+o(1))=\frac{0,5307}{s}(1+o(1))$.
\medskip

\textbf{Замечание 2.}\quad
В частном случае  дизъюнктивных кодов со списочным декодированием при объеме списка~$L=1$
 нижняя граница~\req{ranL1}-\req{ranL3} и асимптотика~\req{ranL-4} совпадают
с нижней границей~\req{ran1}-\req{ran3} и асимптотикой~\req{ran4}.

\textbf{Замечание 3.} \quad
В работе~\cite{chd08} приводится основанная, к сожалению,  на ошибочных рассуждениях более сильная, чем правая часть~\req{ranL-5},
 нижняя граница для скорости~\req{limR_L}.

Правая часть~\req{ranL-5} дает наилучшую известную к
настоящему времени нижнюю границу для максимальной скорости~\req{limR_L} двухступенчатых групповых проверок.
Остается открытой задача получения верхней границы для скорости~\req{limR_L}, уточняющей
очевидную верхнюю границу ${R}_{\infty}(s)\le 1/s$, которая вытекает из~\req{lb-triv1}.

В таблице 2 представлены численные значения нижней границы
скорости СД $\;s_L$-кодов при малых значениях
параметров $s$ и $L$
и указана соответствующая доля $Q_L(s)$ оптимального веса кодовых слов для ансамбля равновесных двоичных кодов
в границе случайного кодирования $\underline{R}_L(s)$ из теоремы~7.
Даны также некоторые численные значения нижней границы~\req{ranL-5}.
При доказательстве утверждений~2 и~3 теоремы~7 для  $Q_L(s)$ будут установлены асимптотические равенства:
\beq{QLs}
Q_L(s) = \frac{\ln 2}{s} + \frac{L \ln^2 2}{s^2} + o \( \frac{1}{s^2} \),\quad s\to\infty,\quad L=1,2,\dots,
\eeq
\beq{QsL}
Q_L(s) = \[ \frac{s^s}{(s-1)^{s-1}} + 1 \]^{-1} + o(1),
\quad L\to\infty,\quad s=2,3,\dots.
\eeq

\begin{table}[ht]
\caption{}\label{tab2}
\begin{tabular*}{\textwidth}{@{\extracolsep{\fill}}cccccc}
\hline
$(s, L)$  & $(2, 2)$ & $(2, 3)$ & $(2, 4)$ & $(2, 5)$ & $(2, 6)$ \cr
$\underline{R}_L(s)$  & $2,35 \cdot 10^{-1}$ & $2,59 \cdot 10^{-1}$ & $2,72 \cdot 10^{-1}$ & $2,81 \cdot 10^{-1}$ & $2,87 \cdot 10^{-1}$ \cr
$Q_L(s)$  & $0,24$ & $0,23$ & $0,23$ & $0,22$ & $0,22$ \cr
\hline
$(s, L)$  & $(3, 2)$ & $(3, 3)$ & $(3, 4)$ & $(3, 5)$ & $(3, 6)$ \cr
$\underline{R}_L(s)$  & $1,14 \cdot 10^{-1}$ & $1,34 \cdot 10^{-1}$ & $1,46 \cdot 10^{-1}$ & $1,55 \cdot 10^{-1}$ & $1,61 \cdot 10^{-1}$ \cr
$Q_L(s)$  & $0,18$ & $0,17$ & $0,16$ & $0,16$ & $0,15$ \cr
\hline
$(s, L)$  & $(4, 2)$ & $(4, 3)$ & $(4, 4)$ & $(4, 5)$ & $(4, 6)$ \cr
$\underline{R}_L(s)$  & $6,84 \cdot 10^{-2}$ & $8,37 \cdot 10^{-2}$ & $9,40 \cdot 10^{-2}$ & $1,01 \cdot 10^{-1}$ & $1,06 \cdot 10^{-1}$ \cr
$Q_L(s)$  & $0,14$ & $0,13$ & $0,13$ & $0,12$ & $0,12$ \cr
\hline
$(s, L)$  & $(5, 2)$ & $(5, 3)$ & $(5, 4)$ & $(5, 5)$ & $(5, 6)$ \cr
$\underline{R}_L(s)$  & $4,55 \cdot 10^{-2}$ & $5,74 \cdot 10^{-2}$ & $6,59 \cdot 10^{-2}$ & $7,22 \cdot 10^{-2}$ & $7,71 \cdot 10^{-2}$ \cr
$Q_L(s)$  & $0.12$ & $0.11$ & $0.11$ & $0.10$ & $0.10$ \cr
\hline
$(s, L)$  & $(6, 2)$ & $(6, 3)$ & $(6, 4)$ & $(6, 5)$ & $(6, 6)$ \cr
$\underline{R}_L(s)$  & $3,25 \cdot 10^{-2}$ & $4,20 \cdot 10^{-2}$ & $4,90 \cdot 10^{-2}$ & $5,44 \cdot 10^{-2}$ & $5,86 \cdot 10^{-2}$ \cr
$Q_L(s)$  & $0,10$ & $0,09$ & $0,09$ & $0,09$ & $0,09$ \cr
\hline
$s$ & $2$ & $3$ & $4$ & $5$ & $6$ \cr
$\underline{R}_{\infty}(s)$ & $0,322 $ & $0,199$ & $0,145$ & $0,114$ & $0,094$  \cr
\hline
\end{tabular*}
\end{table}

\subsection{Дизъюнктивные планы поиска}
\quad
\textbf{Определение 3.}~\cite{ks64},\cite{dr83}.\quad
Код $X$ называется {\em дизъюнктивным $s$-планом} ($(\le s)$-{\em планом}),
если  дизъюнктивная (булева) сумма любого набора, содержащего ровно  $s$ ($\le s$)
столбцов кода~$X$, отлична от дизъюнктивной  суммы  любого другого набора,
содержащего ровно $s$  ($\le s$) столбцов кода~$X$.
Обозначим через $N(t,=s)$ ($N(t,\le s)$) минимальное число строк дизъюнктивного $s$-плана ($(\le s)$-плана)
объема $t$ и определим {\em скорость} дизъюнктивных $s$-планов ($(\le s)$-планов)
 \beq{Rdes}
 R(=s)\eq\varlimsup_{t\to\infty}\frac{\log_2 t}{N(t,=s)},\qquad
 R(\le s)\eq\varlimsup_{t\to\infty}\frac{\log_2 t}{N(t,\le s)} .
 \eeq

Очевидно~\cite{ks64}, что скорость
\beq{Rdes-triv}
R(\le s)\,\le\, R(= s)\le 1/s,\qquad s=1,2,\dots,
\eeq
а определение 3 дает  необходимое и достаточное условие
однозначного восстановления  при планировании $N$ неадаптивных групповых проверок,
описываемых строками кода $X$, для дизъюнктивной модели поиска $s$ $\;(\le s)$ дефектных
элементов во множестве из  $t$ элементов.
Семейство из $t$ подмножеств множества $[N]$ , для которого дизъюнктивный $s$-план
является матрицей инцидентности,  называется  семейством,
{\em  свободным от объединений}~\cite{cop98}.
Для скоростей~\req{Rsl}-\req{RsL} и~\req{Rdes} справедливы неравенства
\beq{Rdes-b}
R(s,1)\,\le R(\le s)\,\le\,R(s-1,1), \qquad
 R(=s)\,\le\, R_2(s-1),\qquad s=2,3,\dots.
 \eeq
Первые два неравенства были отмечены в~\cite{ks64}, а третье было
установлено  в~\cite{dr83} как следствие предложения~3.

Вычисления по формулам~\req{recL1}-\req{recL4} из теоремы~6  дают:  $s(1)=2$, $s(2)=6$, $s(3)=12$,
$s(4)=20$, $s(5)=25$, $s(6)=36$,\ldots.\quad При  $L=2$
и $s=7,8\dots,14$\quad
получаем следующие значения верхней границы  $\overline{R}_2(s-1)$:
\begin{center}
\begin{tabular}{|c||c|c|c|c|c|c|c|c|}
\hline
$s$ & 7 & 8 & 9 & 10 & 11 & 12 & 13 & 14\\
\hline
$1/s$ & 0,143 & 0,125 & 0,111 & 0,100 & 0,091 & 0,083 & 0,077 & 0,071\\
\hline
$\overline{R}_2(s-1)$ & 0,163 & 0,141 & 0,117 & 0,102 & 0,086
& 0,076 & 0,066 & 0,059\\
\hline
\end{tabular}
\end{center}
\centerline{Таблица 3.}
Таблица~3 показывает, что верхняя граница $\overline{R}_2(s-1)<1/s$, если $s\ge 11$. Поэтому
неравенство~\req{Rdes-b} означает, что скорость дизъюнктивных  $s$-планов
$R(=s)<1/s$ при $s\ge 11$. Для $s=2$ нетривиальное неравенство $R(=2)\le 0,4998<1/2$
было доказано в работе~\cite{cop98}.
При $3\le s\le10$ неравенство  $R(=s)<1/s$ можно рассматривать как нашу гипотезу.


\section{Доказательства теорем}
\subsection{Доказательство теоремы 1}
\quad
Пусть при фиксированном  $s=2,3,\dots$ функция  $f_s(x)$ параметра $x$,  $0<x<1$,
определена~\req{entropy}-\req{rec1}, а
$$
K_s\,\eq\,\left[\max_{0\leq x\leq\frac{K_s-K_{s-1}}{K_s}}f_s(x)\right]^{-1}, \quad s=2,3,\dots ,\quad K_1 \eq 1,
$$
обозначает рекуррентную последовательность, введенную в~\cite{dr82}. Тогда утверждение
теоремы 1, т.е. неравенство~\req{low-non-rec},
 равносильно неравенству
\beq{upKs}
K_s\,\le\,\frac{(s+1)^2}{2\log_2[(s+1)/8]},\qquad s\ge8.
\eeq

При $9\leq s\leq 236$  справедливость~\req{upKs} проверяется с помощью компьютерного счета.
При $s\ge237$ доказательство~\req{upKs} опирается на следующие свойства
последовательности $K_s$:
\beq{K1}
K_s = \left[f_s\left(1 - \frac{K_{s-1}}{K_s}\right)\right]^{-1},\quad
1-\frac{K_{s - 1}}{K_s}\,<\, v_s,
\quad s\ge3;
\eeq
\beq{K2}
v_s\,> \,\frac{s}{1+se}\,> \,\frac{2}{s},\qquad s\ge8,
\eeq
в формулировках которых использованы~\req{entropy}-\req{rec2}.
Для фиксированного $s\ge3$ соотношения~\req{K1} и первое неравенство в~\req{K2}
были установлены в~\cite{dr82}-\cite{dmtv02}. Для $s\ge8$ второе неравенство в~\req{K2} очевидно.
Кроме того, при $0<x<1$ и $s\ge2$ имеют место оценки
\begin{multline}\label{low-f}
f_s(x) \eq -\frac{x}{s}\left(\log_2 x - \log_2 s\right) - \left(1 - \frac{x}{s}\right) \log_2\left[1-\frac{x}{s}\right] - \frac{x}{s}\log_2 s + x\left(1-\frac{1}{s}\right)\log_2\left[1-\frac{1}{s}\right]=\\
=-\frac{x}{s}\log_2 x + x\left(1-\frac{1}{s}\right)\log_2\left[1-\frac{1}{s}\right] - \left(1-\frac{x}{s}\right)\log_2\left[1-\frac{x}{s}\right]\geq\\
\geq-\frac{x}{s}\log_2 x + x\left(1-\frac{1}{s}\right)\log_2\left[1-\frac{1}{s}\right] + \left(1-\frac{x}{s}\right)\frac{x}{s}\log_2 e,
\quad 0<x<1,\; s\ge2,
\end{multline}
и
\beq{log2e}
\left(s-1\right)\log_2\left[1-\frac{1}{s}\right]>-\log_2 e,\quad s\ge2,
\eeq
которые вытекают из определений~\req{entropy}-\req{rec1} функции $f_s(x)$  и
стандартного логарифмического неравенства
\beq{ln}
\ln u\,\le\, u-1, \qquad u>0.
\eeq

Далее при выводе~\req{upKs} для  $s\ge237$  анализируются отдельно  два случая.
 В первом случае  рассматриваются  значения $s\ge237$, когда
$1 - \frac{K_{s-1}}{K_s}>\frac{2}{s}$, а во втором случае -- значения
$s\ge237$, когда $1 - \frac{K_{s-1}}{K_s}\le\frac{2}{s}$.
\begin{enumerate}
\item
Пусть $s\ge237$ и $1 - \frac{K_{s-1}}{K_s}>\frac{2}{s}$.  Тогда,
применяя свойство монотонного возрастания функции $f_s(x)$, $0<x\le v_s$,
и~\req{K1}-\req{log2e},
 несложно проверить  цепочку утверждений:
\begin{multline*}
K_s=\frac{1}{f_s\left(1-\frac{K_{s - 1}}{K_s}\right)}<\frac{1}{{f_s}\left(\frac{2}{s}\right)}
\leq\frac{1}{\frac{2}{s^2}\log_2\left[\frac{s}{2}\right]+\frac{2\log_2 e}{s^2}\left(1-\frac{2}{s^2}\right)+\frac{2(s-1)}{s^2}\log_2\left[1-\frac{1}{s}\right]}=\\
=\frac{s^2}{2\log_2\left[\frac{s}{2}\right]+2\log_2 e-\frac{4\log_2 e}{s^2}+2(s-1)\log_2\left[1-\frac{1}{s}\right]}<\frac{s^2}{2\log_2\left[\frac{s}{4}\right]}.
\end{multline*}
Таким образом, справедливость~\req{upKs} для данного случая доказана.
\item
Пусть $s\ge237$ и $1 - \frac{K_{s-1}}{K_s}\leq\frac{2}{s}$. Определим величину $t_s\eq 1-\frac{K_{s-1}}{K_s}\le\frac{2}{s}$.
Поскольку справедливы оценки~\req{low-f}-\req{log2e}, имеем
\begin{multline} \label{axin}
f_s(x)\ge\frac{x}{s}\(-\log_2x + \log_2e-\frac{x\log_2e}{s}+(s-1)\log_2\left[1-\frac{1}{s}\]\)\geq \\
\geq \frac{x}{s}\(-\frac{x\log_2e}{s}-\log_2 x\), \quad 0<x<1,\; s\ge2.
\end{multline}
Заметим, что функция $q_s(x)\eq\(-\frac{x\log_2e}{s}-\log_2 x\)$ монотонно убывает при $x>0$. Следовательно, $q_s(t_s)\geq q_s\left(\frac{2}{s}\right)$,
потому что $t_s\leq\frac{2}{s}$. В силу неравенства~\req{axin} это означает
\beq{K4}
f_s(t_s)\ge \frac{t_s}{s}q_s(t_s)\ge \frac{t_s}{s}q_s\left(\frac{2}{s}\right)= \frac{t_s}{s}\(-\frac{2\log_2 e}{s^2}+\log_2\left[\frac{s}{2}\right]\)>0.
\eeq

Таким образом, можем написать
\beq{K5}
K_s - K_{s-1}=K_s t_s=\frac{t_s}{f_s(t_s)}
\leq\frac{s}{\log_2\left[\frac{s}{2}\right]-\frac{2\log_2 e}{s^2}}
\leq \frac{s}{\log_2\left[\frac{s}{4}\right]}, \quad s\ge8,
\eeq
где в первых двух равенствах воспользовались видом $t_s$ и соотношением~\req{K1}, затем применили~\req{K4} и наконец учли, что $s\geq 8$.
Другими словами, мы установили рекуррентное неравенство
$$
K_s\,\le\,K_{s-1}\,+\, \frac{s}{\log_2\left[\frac{s}{4}\right]}, \quad s\ge8,
$$
 которое для рассматриваемого случая дает~\req{upKs} при $s\ge237$,
если показать, что
\beq{K6}
\frac{s^2}{2\log_2\left[\frac{s}{8}\right]}+\frac{s}{\log_2\left[\frac{s}{4}\right]}<
\frac{(s+1)^2}{2\log_2\left[\frac{s+1}{8}\right]}, \quad s\ge237.
\eeq
В силу логарифмического свойства~\req{ln}, неравенство~\req{K6} вытекает из
\beq{K7}
\frac{s^2}{2\log_2\left[\frac{s}{8}\right]}+\frac{s}{\log_2\left[\frac{s}{4}\right]}<
\frac{(s+1)^2}{2\(\log_2s + \frac{\log_2e}{s} - 3\)}, \quad s\ge237.
\eeq
Элементарными преобразованиями проверяем, что~\req{K7} эквивалентно неравенству
$$
s\left((2-\log_2 e)\log_2 s + 2\log_2 e - 6\right) + \log_2^2 s - (5+2\log_2 e)\log_2 s + 6 + 6\log_2 e > 0,
$$
справедливость которого при $s \ge 237$ подтверждается очевидным образом.
\end{enumerate}

Теорема 1 доказана.$\quad \square$

\subsection{Выводы границ скорости СП $(s,\l)$-кодов}
\subsubsection{Доказательство теоремы 4}
\quad
\textbf{Доказательство утверждения 1.}\quad
При выводе теоремы~4, как и ниже при выводе теоремы~7, применяем метод случайного кодирования
для ансамбля двоичных равновесных кодов, который является обобщением метода,
разработанного нами  в~\cite{drr89}  для классического случая
дизъюнктивных $s$-кодов. Зафиксируем параметр $Q$, $0<Q<1$. Будем использовать обозначения~\req{X}-\req{Rsl}, введенные для определения СП $(s,\l)$-кода $X$ длины $N$ и объема~$t$.  Для произвольного кода $X$ и произвольного множества $\S\subset[t]$,
символом $\x(\S)\eq\{\,\x(j)\,:\,j\in\S\}$ обозначим соответствующее подмножество кодовых слов кода~$X$. Для любых непересекающихся множеств
 $\S,\,\L\subset[t]$, $|\S|=\s$, $|\L|=\ell$, $\S\cap\L=\varnothing$,
соответствующую пару $\left(\x(\S),\x(\L)\right)$ подмножеств кодовых слов
кода~$X$ будем называть $(s,\ell)$-{\em хорошей парой}, если существует строка $\x_i$, $i\in [N]$, в которой
$$
 x_i(j)=0 \quad \text{для любого}\quad j\in\S,\quad \text{и}\quad  x_i(k)=1\quad \text{для любого}\quad k\in\L.
$$
В противном случае, пару $\left(\x(\S),\x(\L)\right)$ будем называть $(s,\ell)$-{\em плохой парой}.
Столбец $\x(j)$ назовем $(s,\ell)$-{\em плохим столбцом  в коде} $X$, если в $X$
найдется $(s,\ell)$-плохая пара  $\left(\x(\S),\x(\L)\right)$  и столбец~$\x(j)\in\x(\L)$.

Определим $E(N,t,Q)$ -- ансамбль двоичных $(N\times t)$-матриц $X$  с
$N$  строками и $t$ столбцами, где столбцы выбираются независимо и равновероятно
из множества, состоящего из $N\choose \lfloor QN\rfloor$ столбцов фиксированного веса $\lfloor QN\rfloor$. Пусть множества $\S$ и $\L$ зафиксированы.
Для ансамбля   $E(N,t,Q)$  символом $P_0(N,Q,s,\ell)$ обозначим  вероятность события:
<<пара $\left(\x(\S),\x(\L)\right)$ является $(s,\ell)$-плохой>>.
Очевидно, что $P_0(N,Q,s,\ell)$ не зависит от выбора~$\S$ и $\L$. Для ансамбля   $E(N,t,Q)$
символом $P_1(N,t,Q,s,\ell)$ обозначим
не зависящую от $j\in[t]$ вероятность события: <<столбец $\x(j)$ является $(s,\ell)$-плохим в коде~$X$>>.
Нетрудно видеть, что
$$
P_1(N,t,Q,s,\ell)\le {{t-1} \choose {s+\ell-1}} {{s+\ell-1} \choose {s}}
P_0(N,Q,s,\ell)\,\le\,\frac{t^{s+\ell-1}}{s!(\ell-1)!}\,P_0(N,Q,s,\ell).
$$
Отсюда вытекает, что   математическое ожидание случайной величины,  определяемой как: <<число  $(s,\ell)$-плохих столбцов в коде~$X$>>,
не превосходит
$$
t\cdot\,P_1(N,t,Q,s,\ell)\,<\,t\, \frac{t^{s+\ell-1}}{s!(\ell-1)!}\,P_0(N,Q,s,\ell).
$$
Поэтому при
$$
t\,<\,\left[\frac{s!(\ell-1)!}{2\,P_0(N,Q,s,\ell)}\right]^{1/(s+\ell-1)}
$$
существует $(N\times t/2)$-матрица $X$, которая является
 СП $(s,\ell)$-кодом. Следовательно, для любого $Q$, $0<Q<1$, максимальный объем
 СП $(s,\ell)$-кодов
$$
t_{cf}(N,s,\ell)\,\ge\,\left\lfloor\,\frac12\,\left[\frac{s!(\ell-1)!}
{2\,P_0(N,Q,s,\ell)}\right]^{1/(s+\ell-1)}\right\rfloor, \quad 0<Q<1.
$$
Тогда, согласно определению~\req{Rsl} скорости
$R(s,\ell)$,  приходим к неравенству
\beq{First}
R(s,\ell)\ge
\underline{R}(s,\l)\eq
\frac{1}{s+\ell-1}\,\max_{0<Q<1}A(s,\ell,Q), \qquad 2\le\ell\le s,
$$
$$
A(s,\ell,Q)\eq\overline{\lim_{N \to \infty}}\frac{-\log_2 P_0(N,Q,s,\ell)}{N}, \quad 0<Q<1.
\eeq
Для завершения вывода утверждения 1 теоремы 4 остается вычислить в явном виде функцию
$A(s,\ell,Q)$ и показать, что правая часть~\req{First} задается~\req{Th4}.

Будем использовать терминологию {\em типов}~\cite{ck85} последовательностей.
Рассмотрим $2$ фиксированных набора, состоящих из двоичных  равновесных столбцов длины $N$:
$$
\{\x(1),\x(2),\dots,\x(s)\} \quad \text{и}\quad \{\y(1),\y(2),\dots,\y(\ell)\},
\quad \text{где}\quad \x(i),\,\y(j)\in\{0,1\}^{N}
$$
и  вес столбца $|\x(i)|=|\y(j)|=\lfloor QN\rfloor$ для любых $i\in[s],\,j\in[\ell]$. Первый набор образует
двоичную $(N\times s)$-матрицу $X_s$, а второй набор -- $(N\times \ell)$-матрицу~$Y_{\l}$.
Сопоставим данным матрицам их {\em типы}, т.е. наборы целых
чисел $\{n(\a)\}$, $\a\eq(a_1,a_2,\dots,a_s)\in\{0,1\}^s$, и
$\{m(\b)\}$, $\b\eq(b_1,b_2,\dots,b_{\l})\in\{0,1\}^{\ell}$,
где элемент набора $n(\a)$, $0\le n(\a)\le N$, ($m(\b)$, $0\le m(\b)\le N$)
определяется как {\em количество строк  в матрице $X_s$ $(Y_{\l})$, совпадающих с
$\a$ $(\b)$}. Очевидно, что для любых двоичных матриц $X_s$ и $Y_{\l}$ сумма
$$
\sum\limits_{\a}\,n(\a)\,=\,\sum\limits_{\b}\,m(\b)\,=\,N.
$$

Через $n\left(\0\right)$ $\(m\left(\1\right)\)$ будем обозначать число нулевых (единичных) строк в $X_s$ $\(Y_{\l}\)$.
Заметим, что если $N - n(\0) < m(\1)$, то соответствующая пара
 $(X_s,Y_{\l})$ является $(s,\ell)$-хорошей.
В остальных случаях, число различных пар
матриц $\(X_s,\,Y_{\l}\)$, которым сопоставлены фиксированные  типы $(\{n(\a)\},\{m(\b)\})$,    равно
$\frac{N!}{\prod_{\a}n(\a)!}\frac{N!}{\prod_{\b}m(\b)!}$, а доля
$(s,\ell)$-плохих пар из общего числа составляет $\frac{{
{N-n\left(\0\right)} \choose m\left(\1\right)}}{{ {N} \choose
m\left(\1\right)}}$. Таким образом, вероятность 
\beq{Pr0}
P_0(N,Q,s,\ell)\,=\,
\sum_{\{n(\a)\}}\sum_{\{m(\b)\}}\frac{N!}{\prod_{\a}n(\a)!}\frac{N!}{\prod_{\b}m(\b)!}
\frac{{ {N-n\left(\0\right)} \choose m\left(\1\right)}}{{ {N} \choose m\left(\1\right)}}
 {N \choose \lfloor QN\rfloor}^{-s-\ell},
\eeq
где суммирование идет по всевозможным типам $\{n(\a)\}$ и $\{m(\b)\}$, для которых
\beq{condD}
\begin{cases}
n(\0)+m(\1)\le N;\\
0\le n(\a)\le N;\quad 0\le m(\b)\le N;\\
\sum\limits_{\a}n(\a)=\sum\limits_{\b}m(\b)=N;\\
|\x(i)|=\sum\limits_{\a:\,a_i=1}n(\a)=|\y(j)|=
\sum\limits_{\b:\,y_j=1}m(\b)=\lfloor QN\rfloor\quad\text{для любых } i\in[s],\,j\in[\ell].\\
\end{cases}
\eeq

 Пусть $N \to \infty$ и
 $ n(\a)\eq N[\tau(\a)+o(1)]$,  $m(\b)\eq N[\chi(\b)+o(1)]$,
  где фиксированные распределения вероятностей $\tau\eq\{\tau(\a)\},\,\a\in\{0,1\}^s$ и
 $\chi\eq\{\chi(\b)\},\,\y\in\{0,1\}^{\ell}$, обладают свойствами, индуцированными условиями ~\req{condD}, т.е.
\beq{CondN}
\begin{cases}
\sum\limits_{\a\in\{0,1\}^s} \tau(\a) = 1,\quad \sum\limits_{\b\in\{0,1\}^{\ell}} \chi(\b) = 1,\quad \tau(\0)+\chi(\1)\le1,\\
 \sum\limits_{\a:\,a_i = 1} \tau(\a) = Q,\quad \sum\limits_{\b:\,b_j=1}\chi(\b) = Q\quad\text{для любых }  i\in[s],\,j\in[\ell].\\
 \end{cases}
\eeq
С помощью формулы Стирлинга для типов, соответствующих этим распределениям, находим логарифмическую
асимптотику слагаемого в~\req{Pr0}:
$$
-\log_2\left\{ \frac{N!}{\prod_{\a}n(\a)!}\frac{N!}{\prod_{\b}m(\b)!}
\frac{{ {N-n\left(\0\right)} \choose m\left(\1\right)}}{{ {N} \choose m\left(\1\right)}}
 {N \choose \lfloor QN\rfloor}^{-s-\ell}\right\}=N[F(\tau,\chi,Q)+o(1)],
$$
где
\begin{multline*}
F=F(\tau,\chi,Q)\eq
\sum_{\a} \tau(\a )\log_2\[\tau(\a )\]
+\sum_{\b} \chi(\b )\log_2\[\chi(\b )\]
-(1-\tau\left(\0\right))h\left(\frac{\chi\left(\1\right)}{1-\tau\left(\0\right)}\right)+\\
+(s+\ell)h(Q)+h(\chi\left(\1\right)).
\end{multline*}

Пусть $\tau_Q\eq\{\tau_Q(\a)\}$,  и $\chi_Q\eq\{\chi_Q(\b)\}$ --
распределения со свойствами~\req{CondN}, на которых достигается
минимум функции $F(\tau,\chi, Q)$ для данного $Q$. Тогда главный
член логарифмической асимптотики суммы слагаемых~\req{Pr0} есть
\beq{Second}
A(s,\ell,Q)\eq\overline{\lim_{N \to \infty}}\frac{-\log_2 P_0(N,Q,s,\ell)}{N}=\min\limits_{(\tau,\chi)\in\req{CondN}}F(\tau,\chi,Q)=F(\tau_Q,\chi_Q,Q).
\eeq
Будем искать минимум  функции $F\eq F(\tau,\chi,Q)$ при  ограничениях~\req{CondN}.
Поскольку функция $F$ непрерывна в рассматриваемой  области допустимых значений аргумента $(\tau,\chi)$,
 в том числе и на ее границе, то достаточно
 найти минимум $F$ при условиях~\req{CondN} с исключенными границами.
Запишем соответствующую  задачу минимизации: $F\to \min$.
\begin{align}\label{Pro}
&\text{Основная функция:}\quad\quad F(\tau,\chi,Q): \mathbb{X}\to \mathbb{R}.& \notag\\
&\text{Ограничения:}\,\,
\begin{cases}
 \sum\limits_{\a\in\{0,1\}^s} \tau(\a) = 1;\\
 \sum\limits_{\b\in\{0,1\}^{\ell}} \chi(\b) = 1;\\
 \sum\limits_{\a:\,a_i = 1} \tau(\a) = Q\quad\text{для любого }  i\in[s];\\
 \sum\limits_{\b:\,b_j=1}\chi(\b) = Q\quad \text{для любого }  j\in[\ell].
 \end{cases}&\\
&\text{Область поиска }\mathbb{X}:\,\,
\begin{cases}
0<\tau(\a)<1\quad \text{для любого }  \a\in\{0,1\}^s;\\
0<\chi(\b)<1\quad \text{для любого } \b\in\{0,1\}^{\ell}; \notag\\
\tau(\0)+\chi(\1)<1.
\end{cases}&
\end{align}

Для вычисления точки минимума $(\tau_Q,\chi_Q)$ применим стандартный метод множителей
Лагранжа. Рассмотрим лагранжиан 
\begin{multline*}
\Lambda \eq F(\tau,\chi, Q) + \lambda_0 \left(\sum_{\a} \tau(\a) - 1\right) + \lambda_1 \left(\sum_{\b} \chi(\b) - 1\right)+\\
+\sum_{i = 1}^s \mu_i \left(\sum_{\a:\,a_i = 1} \tau(\a) - Q\right)+\sum_{i = 1}^{\ell} \nu_i \left(\sum_{\b:\,b_i = 1} \chi(\b) - Q\right).
\end{multline*}
Необходимые условия экстремального распределения $(\tau_Q,\chi_Q)$ имеют вид:
\beq {Cond}
\begin{cases}
\frac{\partial \Lambda}{\partial(\tau(\a))} = \log_2 \[\tau(\a)\] + \log_2 e + \lambda_0 + \sum\limits_{i:\,a_i = 1} \mu_i\,=\,0 \quad \text{для любого }  \a \neq \0; \\
\frac{\partial \Lambda}{\partial(\tau(\0)} = \log_2 \[\tau(\0)\] + \log_2 e + \lambda_0 + \log_2\[\frac{1-\tau\left(\0\right)}{1-\tau\left(\0\right)-\chi\(\1\)}\]=0;\\
\frac{\partial \Lambda}{\partial(\chi(\b))}=\log_2\[ \chi(\b)\] + \log_2 e +\lambda_1 + \sum\limits_{i:\,b_i = 1}\nu_i\,=\,0\quad \text{для любого } \b \neq \1; \\
\frac{\partial \Lambda}{\partial(\chi\left(\1\right))}=\log_2 \[\chi\left(\1\right)\] + \log_2 e +\lambda_1 + \sum\limits_{i= 1}^{\ell}\nu_i+\log_2\[\frac{1-\chi\left(\1\right)}{1-\tau\left(\0\right)-\chi\left(\1\right)}\]\,=\,0.\\
\end{cases}
\eeq

Покажем, что матрица, составленная из вторых производных лагранжиана, является положительно определенной.
Действительно, опишем эту матрицу:
\begin{align*}
&\frac{\partial^2 \Lambda}{\partial^2(\tau(\a))}=\frac{\log_2 e}{\tau(\a)}>0\quad \text{для любого }  \a\neq\0;\\
&\frac{\partial^2 \Lambda}{\partial^2(\chi(\b))}=\frac{\log_2 e}{\chi(\b)}>0\quad \text{для любого } \b\neq\1;\\
&a'\eq\frac{\partial^2 \Lambda}{\partial(\tau\left(\0\right))^2}=\frac{\log_2 e}{\tau\left(\0\right)}+\frac{\log_2 e\cdot\chi\left(\1\right)}{(1-\tau\left(\0\right))(1-\tau\left(\0\right)-\chi\left(\1\right))}>0;\\
&b'\eq\frac{\partial^2 \Lambda}{\partial(\tau\left(\0\right))\partial(\chi\left(\1\right))}=\frac{\log_2 e}{1-\tau\left(\0\right)-\chi\left(\1\right)}>0;\\
&c'\eq\frac{\partial^2 \Lambda}{\partial(\chi\left(\1\right))^2}=\frac{\log_2 e}{\chi\left(\1\right)}+\frac{\log_2 e\cdot\tau\left(\0\right)}{(1-\chi\left(\1\right))(1-\tau\left(\0\right)-\chi\left(\1\right))}>0,
\end{align*}
а остальные  элементы нулевые. Поэтому,  достаточно проверить, что~$a'c'-{b'}^2>0$. Имеем
\begin{multline*}
\frac{a'c'-{b'}^2}{(\log_2 e)^2}=\frac{1}{\tau\left(\0\right)\chi\left(\1\right)}+\frac{1}{(1-\tau\left(\0\right))(1-\tau\left(\0\right)-\chi\left(\1\right))}+
\frac{1}{(1-\chi(\left(\1\right))(1-\tau\left(\0\right)-\chi\left(\1\right))}+\\
+\frac{\tau\left(\0\right)\chi\left(\1\right)}{(1-\tau\left(\0\right))(1-\chi\left(\1\right))(1-\tau\left(\0\right)-\chi\left(\1\right))^2}-\frac{1}{(1-\tau\left(\0\right)-\chi\left(\1\right))^2}=\\
=\frac{1}{\tau\left(\0\right)\chi\left(\1\right)}+\frac{1}{(1-\tau\left(\0\right))(1-\tau\left(\0\right)-\chi\left(\1\right))}+
\frac{1}{(1-\chi\left(\1\right))(1-\tau\left(\0\right)-\chi\left(\1\right))}-\\
-\frac{1}{(1-\tau\left(\0\right))(1-\chi\left(\1\right))(1-\tau\left(\0\right)-\chi\left(\1\right))}\ge \frac{1-\tau\left(\0\right)\chi\left(\1\right)}{1-\tau\left(\0\right)-\chi\left(\1\right)}>0.
\end{multline*}
Матрица вторых производных функции $F$ совпадает с вышеописанной матрицей и, следовательно~\cite{opu},
 функция $F$ строго выпукла в области $\mathbb{X}$.

Заметим, что уравнения ограничений~\req{Pro} образуют аффинное подпространство $\mathbb{G}$ в $\mathbb{R}^{2^s+2^{\ell}}$ размерности $\(2^s+2^{\ell}-(s+\ell+2)\)$. Откуда вытекает, что функция $F$ строго выпукла и в $\mathbb{G}\cap \mathbb{X}$,
что в свою очередь означает, что в $\mathbb{G}\cap \mathbb{X}$ локальный минимум функции $F$ является глобальным и единственным.
Далее воспользуемся теоремой Каруша-Куна-Таккера~\cite{opu}, утверждающей, что всякое решение,
удовлетворяющее системе~\req{Cond}, ограничениям~\req{Pro} и имеющее положительно определенную матрицу вторых производных
лагранжиана в этой точке, является локальным минимумом функции $F$.
Таким образом, если  есть решение системы~\req{Cond} и~\req{Pro} в области $\mathbb{X}$, то оно единственно,
и эта точка является минимумом функции $F$ на $\mathbb{X}$.

Докажем, что из симметрии задачи следует равенство: $\mu\eq\mu_1 = \mu_2 = ... = \mu_s $.
Достаточно показать, что $ \mu_i = \mu_j $ для $ i \not= j $.
Пусть $\bar\a_i \eq (0,\dots,1,\dots,0) $ обозначает $s$-строку, в которой на $i$-м месте стоит~$1$.
При перестановке индексов $i$ и $j$ мы получим задачу минимизации, эквивалентную исходной. Следовательно,
если $ (\tau_Q^1,\chi_Q) $ - решение, то решением будет также и  $ (\tau_Q^2,\chi_Q) $, для которого
вероятность $ \tau_Q^2(\a) = \tau_Q^1(\tilde \a) $, где $ \tilde \a $ -- строка, полученная перестановкой индексов
$i$ и $j$ из строки $\a$. Из единственности решения $ \tau_Q$ следует, что распределения
$ (\tau_Q^1,\chi_Q) $ и $ (\tau_Q^2,\chi_Q) $ совпадают.  В частности, вероятность
$ \tau_Q^1(\bar \a_i) = \tau_Q^1(\bar \a_j) $. Равенство множителей Лагранжа вытекает из первого уравнения
в системе~\req{Cond}.
 Используя те же рассуждения, можно доказать, что $\nu\eq\nu_1 = \nu_2 = ... = \nu_{\ell}$.

Для сокращения записи введем параметры $\hat{\mu}\eq \log_2 e + \lambda_0,\, \hat{\nu}\eq\log_2 e + \lambda_1$.
Тогда уравнения~\req{Cond} примут вид
\beq {Cond2}
\begin{cases}
\hat{\mu} + \mu \sum_{i = 1}^s a_i + \log_2\[ \tau(\a)\] = 0  \quad\text{при }\a \neq \0;\\
\hat{\mu} + \log_2\[\tau\left(\0\right) \] +\log_2\[\frac{1-\tau\left(\0\right)}{1-\tau\left(\0\right)-\chi\(\1\)}\]= 0;  \\
\hat{\nu}+\nu\sum_{i = 1}^{\ell} b_i + \log_2\[ \chi(\b)\] = 0  \quad\text{при }\b \neq \1;  \\
\hat{\nu}+\nu\ell+ \log_2 \[ \chi\left(\1\right)\]+\log_2\[\frac{1-\chi\left(\1\right)}{1-\tau\left(\0\right)-\chi\left(\1\right)}\]= 0.
\end{cases}
\eeq
Из первого уравнения системы~\req{Cond2} следует, что вероятность
$$
\tau(\a) = 2^{-\hat{\mu}}2^{-\mu \sum a_i}=\frac{2^{-\hat{\mu}}}{z^s} \prod_{i = 1}^s \tilde {P}_1(a_i)
\quad \text{при }\a \neq \0,
$$
где
$$
\tilde {P}_1(0) \eq \frac{1}{1 + 2^{-\mu}} \eq z, \quad
\tilde {P}_1(1) \eq \frac{2^{-\mu}}{1 + 2^{-\mu}} \eq 1 - z.
$$
Из условий \req{Pro} получаем
$$
Q = \frac{2^{-\hat{\mu}}}{z^s} \sum_{k = 0}^{s-1} {s-1\choose k} z^{s - k - 1}(1-z)^{k + 1} =
\frac{1 - z}{2^{\hat{\mu}}z^s}\quad \Leftrightarrow\quad \hat{\mu} = \log_2\[ \frac{1 - z}{Qz^s}\].
$$
Далее, так как $ \tau=\{\tau(\a)\} $, $\a\in\{0,1\}^s$,
 является распределением вероятностей, то
$$
1 - \tau\left(\0\right) = \sum_{\a > \0} \tau(\a) =
\frac{2^{-\hat{\mu}}}{z^s} \sum_{k = 1}^s {s \choose k} z^{s - k} (1 - z)^k = \frac{Q(1-z^s)}{1 - z}.
$$
Поэтому все вероятности экстремального распределения $\tau_Q=\{\tau_Q(\a)\}$
могут быть представлены как функции независимой переменной $z$, $0<z<1$:
 \beq{z}
\tau_Q(\a) = \frac{Q}{1 - z} z^{s - \sum_{i=1}^s a_i} (1 - z)^{\sum_{i=1}^s a_i}\quad\text{при } \a\neq\0;
\quad \tau_Q\left(\0\right) = 1-\frac{Q(1 - z^s)}{1 - z}.
\eeq
Аналогичным образом, из третьего уравнения системы~\req{Cond2} следует, что вероятность
$$
\chi(\b) = 2^{-\hat{\nu}}2^{-\nu \sum b_i}=\frac{2^{-\hat{\nu}}}{u^\l} \prod_{i = 1}^\l \tilde {P}_2(b_i)
\quad \text{при }\b \neq \1,
$$
где
$$
\tilde {P}_2(0) \eq \frac{1}{1 + 2^{-\nu}} \eq u, \quad
\tilde {P}_2(1) \eq \frac{2^{-\nu}}{1 + 2^{-\nu}} \eq 1 - u.
$$
Далее, так как $ \chi=\{\chi(\b)\} $, $\b\in\{0,1\}^\l$,
является распределением вероятностей, то
$$
1 - \chi\left(\1\right) = \sum_{\b < \1} \chi(\b) =
\frac{2^{-\hat{\nu}}}{u^\l} \sum_{k = 0}^{\l-1} {\l \choose k} u^{\l - k} (1 - u)^k = \frac{2^{-\hat{\nu}}}{u^\l}(1 - (1-u)^\l).
$$
Из условий \req{Pro} и предыдущего уравнения получаем 
\begin{multline*}
Q = \frac{2^{-\hat{\nu}}}{u^\l} \sum_{k = 0}^{\l-2} {\l-1\choose k} u^{\l - k - 1}(1-u)^{k + 1}  + \chi(\1)= \\
= 1 + \frac{2^{-\hat{\nu}}}{u^\l}\left((1-u)(1-(1-u)^{\l-1}) - (1-(1-u)^\l)\right).
\end{multline*}
Откуда имеем
$$
\frac{2^{-\hat{\nu}}}{u^\l} = \frac{1 - Q}{u}\quad\Leftrightarrow\quad \hat{\nu} = \log_2\[ \frac{u}{(1-Q)u^\l}\].
$$
Тогда вероятности экстремального распределения $\chi_Q=\{\chi_Q(\b)\}$,
$\b\in\{0,1\}^{\l}$,
можно представить в виде функций независимой переменной $u$, $0<u<1$:
\beq{w}
\chi_Q(\b) = \frac{1 - Q}{ u} u^{\ell - \sum_{j=1}^{\l} b_j} (1 - u)^{\sum_{j=1}^{\l} b_j}\quad \text{при }\b\neq\1; $$$$
\quad \chi_Q\left(\1\right) = 1-\frac{(1 - Q)(1-(1-u)^{\ell})}{u}.
\eeq

Подставив найденные  выражения для $ \mu,\,\hat{\mu},\,\nu,\,\hat{\nu},\,\tau_Q\left(\0\right)$ и $\chi_Q\left(\1\right) $
во второе уравнения системы~\req{Cond2},  имеем:
\begin{multline*}
\log_2\[\frac{1 - z}{Qz^s}\] + \log_2\[1 - \frac{Q(1-z^s)}{1 - z}\] + \log_2\[\frac{Q(1-z^s)}{1 - z}\]=
\\
= \log_2\[\frac{Q(1-z^s)}{1 - z} - 1 + \frac{(1 - Q)(1-(1-u)^{\ell})}{u}\].
\end{multline*}
Откуда получаем
\begin{equation*}
\frac{(1-z - Q(1-z^s))(1-z^s)}{z^s} = \frac{Q(1-z^s)u - (1-z)u + (1-Q)(1-z)(1-(1-u)^\l)}{u},
\end{equation*}
\begin{multline*}
(1-z)(1-z^s)u+(1-z)uz^s-(1-z)(1-(1-u)^\l)z^s =\\
=Q(u(1-z^s)^2 + (1-z^s)uz^s - (1-z)(1-(1-u)^\l)z^s),
\end{multline*}
\beq{LinkQzu1}
Q = \frac{(1-z)(u - z^s(1-(1-u)^\l))}{((1-z^s)u - z^s(1-z)(1-(1-u)^\l))}.
\eeq
Подставив найденные  выражения для $ \mu,\,\hat{\mu},\,\nu,\,\hat{\nu},\,\tau_Q\left(\0\right)$ и $\chi_Q\left(\1\right) $
в четвертое уравнения системы~\req{Cond2},  имеем:
\begin{multline*}
\log_2\[ \frac{u}{(1-Q)u^\l}\] - \l\log_2\[\frac{1-u}{u}\] +\log_2\[1 - \frac{(1 - Q)(1-(1-u)^{\ell})}{u}\] +
\\
+ \log_2\[\frac{(1 - Q)(1-(1-u)^{\ell})}{u}\] = \log_2\[\frac{Q(1-z^s)}{1 - z} - 1 + \frac{(1 - Q)(1-(1-u)^{\ell})}{u}\].
\end{multline*}
Откуда получаем
\begin{multline*}
\frac{(u - (1-Q)(1-(1-u)^\l))(1-(1-u)^\l)}{(1-u)^\l } =
\\
= \frac{Q(1-z^s)u - (1-z)u + (1-Q)(1-z)(1-(1-u)^\l)}{1-z},
\end{multline*}
\begin{multline*}
(1-z)(u-1 +(1-u)^\l)(1-(1-u)^\l) + u(1-z)(1-u)^\l-(1-z)(1-u)^\l(1-(1-u)^\l) =
\\ 
= Q((1-u)^\l(1-z^s)u-(1-u)^\l(1-z)(1-(1-u)^\l) - (1-z)(1-(1-u)^\l)^2),
\end{multline*}
\beq{LinkQzu2}
Q =\frac{ (1-z)((1-u - (1-u)^\l)}{ ( (1-z)(1-(1-u)^\l) - u(1-u)^\l(1-z^s))}
\eeq
Из уравнений \req{LinkQzu1} и \req{LinkQzu2} следует, что
\begin{multline*}
((1-u - (1-u)^\l)((1-z^s)u - z^s(1-z)(1-(1-u)^\l))
\\ = (u - z^s(1-(1-u)^\l))( (1-z)(1-(1-u)^\l) - u(1-u)^\l(1-z^s)).
\end{multline*}
Далее элементарными преобразованиями можно получить
\beq{StrongLinkZU}
z-u = z^s (1-u)^\l.
\eeq
Используя последнее уравнение, можно упростить выражение \req{LinkQzu1}
\beq{StrongLinkQZU}
Q = \frac{1-z}{1- (z-u)} \quad\Leftrightarrow\quad 1-Q = \frac{u}{1-(z-u)}.
\eeq
Перепишем \req{z}, используя последнее равенство
\beq{zQ}
\tau_Q(\a) = \frac{z^{s - \sum_{i=1}^s a_i} (1 - z)^{\sum_{i=1}^s a_i}}{1 - (z-u)} \quad\text{при } \a\neq\0; $$$$
\tau_Q\left(\0\right) = 1-\frac{(1 - z^s)}{1 - (z-u)} = \frac{z^s - (z-u)}{1-(z-u)}.
\eeq
Также перепишем \req{w}, применяя \req{StrongLinkQZU}
\beq{wQ}
\chi_Q(\b) = \frac{u^{\ell - \sum_{j=1}^{\l} b_j} (1 - u)^{\sum_{j=1}^{\l} b_j}}{1-(z -u)} \quad \text{при }\b\neq\1; $$$$ \chi_Q\left(\1\right) = 1-\frac{(1-(1-u)^{\ell})}{1-(z -u)} = \frac{(1-u)^\l - (z-u)}{1-(z-u)}.
\eeq

Заметим, что любое решение $0< z,u < 1$  уравнения~\req{StrongLinkZU} задает  распределения  $\tau_Q$ и~$\chi_Q$. В силу единственности такого решения для фиксированного $Q$, то для вычисления  максимума в~\req{Second} по параметру $Q$, $0<Q<1$,
можно применять  равенства:
\beq{Q-z-1}
\max_{0<Q<1}A(s,\ell,Q)\,=\,
\max\limits_{0<Q<1}F\(\tau_Q,\chi_Q,Q\)=\max\limits_{\substack{\req{StrongLinkZU}\\ 0< z,u < 1}}F\(\tau_{Q(z,u)},\chi_{Q(z, u)},\,Q(z,u)\),\\
\eeq
где
\begin{multline}\label{ForR}
F\(\tau_{Q},\chi_{Q}\,Q\)=
\sum_{\a}\tau_Q(\a)\log_2\[\tau_Q(\a)\]+\sum_{\b} \chi_Q(\b)\log_2\[\chi_Q(\b)\]-
\\
-(1-\tau_Q\left(\0\right)) h\left(\frac{\chi_Q\left(\1\right)}{1-\tau_Q\left(\0\right)}\right)
+(s+\ell)h(Q)+h(\chi_Q\left(\1\right)).
\end{multline}
Используя~\req{zQ}-\req{wQ},
представим  все пять слагаемых  суммы в правой части~\req{ForR}
в виде функций переменных $z$ и $u$, связанных между собой посредством \req{StrongLinkZU}. Запишем в таком виде первое слагаемое:
\begin{multline}\label{F1}
\sum_{\a}\tau_Q(\a)\log_2\[\tau_Q(\a)\]=
\left\{\sum_{k=1}^{s}{s\choose k}\frac{z^{s-k}(1-z)^k}{1-(z-u)}
\log_2\[\frac{z^{s-k}(1-z)^{k}}{1-(z-u)}\]\right\}+
\\
+\frac{z^s - (z-u)}{1-(z-u)}
\log_2\[\frac{z^s - (z-u)}{1-(z-u)}\]= \left\{\frac{s(z-z^s)}{1-(z-u)}\log_2 z+\right.
\\
\left.+\frac{s(1-z)}{1-(z-u)}\log_2[1-z]
-\frac{1-z^s}{1-(z-u)}\log_2\[1-(z-u)\]\right\}+
\\
+\frac{z^s - (z-u)}{1-(z-u)}\log_2\[z^s - (z-u)\] - \frac{z^s - (z-u)}{1-(z-u)}
\log_2\[1-(z-u)\]=
\\
=\frac{s(z-z^s)}{1-(z-u)}\log_2 z+\frac{s(1-z)}{1-(z-u)}\log_2[1-z]+\\
+\frac{z^s - (z-u)}{1-(z-u)}\log_2\[z^s - (z-u)\] -\log_2\[1-(z-u)\] = \\
= \frac{su}{1-(z-u)}\log_2 z +\frac{z^s - (z-u)}{1-(z-u)}\log_2\[1-(1-u)^\l\] + \\
+ \frac{s(1-z)}{1-(z-u)}\log_2[1-z]-\log_2\[1-(z-u)\].
\end{multline}
Аналогично второе слагаемое:
\begin{multline}\label{F2}
\sum_{\b} \chi_Q(\b)\log_2\[\chi_Q(\b)\]=\sum_{k=0}^{\ell - 1}{\ell\choose k}\frac{u^{\ell-k}(1-u)^k}{1-(z-u)}
\log_2\[\frac{u^{\ell-k}(1-u)^{k}}{1-(z-u)}\]+
\\
+\frac{(1-u)^\l - (z-u)}{1-(z-u)}\log_2\[\frac{(1-u)^\l - (z-u)}{1-(z-u)}\]= 
\\
=\frac{\ell u}{1-(z-u)}\log_2 u 
+\frac{\ell\left((1-u)-(1-u)^{\ell}\right)}{1-(z-u)}\log_2[1-u]+ 
\\
+\frac{(1-u)^\l - (z-u)}{1-(z-u)}\log_2\[(1-u)^\l - (z-u)\]
-\log_2\[1-(z-u)\] =
\\
=\frac{\ell u}{1-(z-u)}\log_2 u + \frac{\ell(1-z)}{1-(z-u)}\log_2[1-u] + 
\\
+\frac{(1-u)^\l - (z-u)}{1-(z-u)}\log_2\[1-z^s\]-\log_2\[1-(z-u)\].
\end{multline}
Третье слагаемое:
\begin{multline}\label{F3}
-(1-\tau_Q\left(\0\right))h\left(\frac{\chi_Q\left(\1\right)}{1-\tau_Q\left(\0\right)}\right)=
-\(1-\tau_Q(\0)\)\log_2\[1-\tau_Q(\0)\]+
\\
+\chi_Q(\1)\log_2\[\chi_Q(\1)\]
+\(1-\tau_Q(\0)-\chi_Q(\1)\)\log_2\[1-\tau_Q(\0)-\chi_Q(\1)\]=
\\
= - \frac{(1 - z^s)}{1 - (z-u)}\log_2\[\frac{(1 - z^s)}{1 - (z-u)}\] + \frac{(1-u)^\l - (z-u)}{1-(z-u)} \log_2\[\frac{(1-u)^\l - (z-u)}{1-(z-u)}\]
\\
+\frac{1 + (z-u)- z^s - (1-u)^\l}{1 - (z-u)}\log_2\[\frac{1 + (z-u)- z^s - (1-u)^\l}{1 - (z-u)}\] = \\
=- \frac{(1 - z^s)}{1 - (z-u)}\log_2\[1 - z^s\]+ \frac{(1-u)^\l - (z-u)}{1-(z-u)} \log_2\[(1-u)^\l - (z-u)\] +
\\
+\frac{1 + (z-u)- z^s - (1-u)^\l}{1 - (z-u)}\log_2\[1 + (z-u)- z^s - (1-u)^\l\] =
\\
=
\frac{1 + (z-u)- z^s - (1-u)^\l}{1 - (z-u)}\log_2\[1 - (1-u)^\l\] +\\ 
+\frac{\l((1-u)^\l - (z-u))}{1-(z-u)} \log_2\[1-u\]
.
\end{multline}
Четвертое слагаемое:
\begin{multline}\label{F4}
(s+\ell)h(Q)=-\frac{(s+\l)(1-z)}{1- (z-u)}\log_2\[\frac{1-z}{1- (z-u)}\]-\frac{(s+\l)u}{1- (z-u)}\log_2\[\frac{u}{1- (z-u)}\]
\\
= (s+\l)\log_2\[1- (z-u)\] - \frac{(s+\l)(1-z)}{1- (z-u)}\log_2\[1-z\] -
\\
-\frac{(s+\l)u}{1- (z-u)}\log_2 u.
\end{multline}
И, наконец, последнее слагаемое:
\begin{multline}\label{F5}
h(\chi_Q\left(\1\right))=-\frac{(1-u)^\l - (z-u)}{1-(z-u)} \log_2\[\frac{(1-u)^\l - (z-u)}{1-(z-u)}\] -
\\
- \frac{(1-(1-u)^{\ell})}{1-(z -u)}\log_2\[\frac{(1-(1-u)^{\ell})}{1-(z -u)}\] = \\
= -\frac{(1-u)^\l - (z-u)}{1-(z-u)} \log_2\[(1-u)^\l - (z-u)\] + \log_2\[1-(z -u)\] -
\\
-\frac{(1-(1-u)^{\ell})}{1-(z -u)}\log_2\[1-(1-u)^{\ell}\] = -\frac{\l((1-u)^\l - (z-u))}{1-(z-u)}\log_2\[1-u\] -
\\
-\frac{(1-u)^\l - (z-u)}{1-(z-u)}\log_2\[1-z^s\]+ \log_2\[1-(z -u)\]-
\\
-\frac{(1-(1-u)^{\ell})}{1-(z -u)}\log_2\[1-(1-u)^{\ell}\] .
\end{multline}
Подстановка~\req{F1}-\req{F5} в~\req{ForR} и группировка подобных слагаемых дает
\beq{Q-z-2}
F\(\tau_{Q(z,u)},\chi_{Q(z,u)},\,Q(z,u)\)=T(z,u, s,\ell),\quad 0<z,u<1, \quad 2\le\ell\le s,
\eeq
где функция $T(z,u,s,\ell)$ определена следующим образом:
\begin{multline}\label{FinalSum}
T(z,u,s,\ell)\eq \frac{su}{1-(z-u)}\log_2 z- \frac{\l(1-z)}{1- (z-u)}\log_2\[1-z\] -
\\
-\frac{s u}{1- (z-u)}\log_2 u+\frac{\ell\left(1-z\right)}{1-(z-u)}\log_2[1-u] + (s+\l -1 )\log_2\[1- (z-u)\] = \\ =\frac{su}{1-(z-u)}\log_2\[\frac{z}{u}\] + \frac{\l(1-z)}{1- (z-u)}\log_2\[\frac{1-u}{1-z}\] +\\
+ (s+\l -1 )\log_2\[1- (z-u)\].
\end{multline}
Поэтому из построения~\req{First} границы случайного кодирования $\underline{R}(s,\ell)$,
равенств~\req{Q-z-1} и формулы~\req{Q-z-2} следует, что скорость
СП $(s,\ell)$-кодов
\beq{RT}
R(s,\ell)\,\ge\,\underline{R}(s,\ell)\eq\frac{1}{s+\ell-1}\max\limits_{\substack{\req{StrongLinkZU} \\ 0<z,\,u<1 }}\,T(z,u, s,\ell).
\eeq
Таким образом, утверждение~1  доказано.
\medskip

Перейдем к доказательству утверждения 2.
Пусть $\ell\ge2$ фиксировано и $s\to\infty$.
Заменим в правой части~\req{RT}  максимум
по $\{z,u\},\,0<z,u<1,$ функции~\req{T} на ее значение при $z=z'\eq 1 - \frac{c}{s}$, где $c$ -- некоторая константа. Значение $u=u'$ возьмем из уравнения
\beq{StrLZU}
z-u = {z}^s (1-u)^\l.
\eeq
Далее для простоты  вместо $z'$ и $u'$ будем писать $z$ и $u$. Легко проверить, что
\beq{Usl}
u =  1 - \frac{c}{s} - \frac{e^{-c}c^\l}{s^\l}\(1 -\frac{c^2}{2s} + o\(\frac{1}{s}\)\),
\eeq
\beq{zmu}
z-u = \frac{e^{-c}c^\l}{s^\l}\(1 -\frac{c^2}{2s} + o\(\frac{1}{s}\)\),
\eeq
\beq{ratzu}
\frac{z}{u} = 1 + \frac{e^{-c}c^\l}{s^\l}\(1 -\frac{c^2 - 2c}{2s} + o\(\frac{1}{s}\)\),
\eeq
\beq{rat1u1z}
\frac{1-u}{1-z} = 1 + \frac{e^{-c}c^{\l-1}}{s^{\l-1}}(1+o(1)).
\eeq
Непосредственно из 
определения~\req{FinalSum} будем искать асимптотику каждого из слагаемого с точностью до $o(1/s^\l)$. В процессе приведения будем  использовать \req{StrLZU}-\req{rat1u1z}. Итак
\begin{multline*}
T(z,u, s,\ell) = \log_2e\(s\(1-\frac{c}{s}\)\frac{e^{-c}c^\l}{s^\l}\(1 -\frac{c^2 - 2c}{2s}\) + \frac{\l e^{-c}c^{\l}}{s^{\l}}\right. -
\\
\left.- s\(1+\frac{\l-1}{s}\)\frac{e^{-c}c^\l}{s^\l}\(1 -\frac{c^2}{2s}\)\) + o\(\frac{1}{s^{\l+1}}\) = \frac{\log_2 e}{e^c}\frac{ c^{\l}}{s^\l}(1+o(1)).
\end{multline*}
Легко видеть, что максимум по $x$ функции $x^{\l}/e^{x}$ достигается при $x=\l$. Откуда получаем
$$
T(z',u', s,\ell) =\frac{\log_2 e}{e^\l}\frac{ \l^{\l}}{s^\l}(1+o(1)),
$$
и
\begin{equation}\label{Rls}
\underline{R}(s,\ell)\,\ge\,
\frac{1}{s+\ell-1}T\(z',u',s,\ell\)= \frac{\log_2 e}{e^\l}\frac{ \l^{\l}}{s^\l}(1+o(1)),
\quad s\to\infty, \;\ell\ge2.
\end{equation}

Для вычисления асимптотики~\req{Qls} доли оптимального веса $Q(s,\ell)$
подставим в формулу~\req{StrongLinkQZU} значение $z=1 -\frac{\l}{s}$ и получим
$$
Q(s,\ell)=\frac{1-z}{1-z^s(1-z)^{\ell}}=\frac{\ell}{s}(1+o(1)),\quad s\to\infty.
$$
Таким образом, утверждение 2 доказано.

\subsubsection{Доказательство теоремы 3}
\quad
Пусть $s\ge\ell\ge2$. Если для фиксированного $p$, $0<p<1$, произведение $sp$ -- целое число,
то положив в правой части неравенства~\req{eng} параметр  $j\eq\l-1$,
получим
$$
R(s,\l)\,\le\,R(s(1-p),1)\cdot\,\frac{(ps)^{ps}\cdot (\l-1)^{\l-1}}{(ps+\l-1)^{ps+\l-1}}.
$$
Если $s\to\infty$ и $\l\ge2$ фиксировано, то применяя для
скорости $R(s(1-p),1)$ асимптотическое равенство~\req{rec-as},
можем написать
$$
R(s,\l)\,\le\,\min_{0<p<1}\;
\left\{\frac{2\log_2[s(1-p)]}{s^2(1-p)^2}
\cdot\,
\frac{(ps)^{ps}\cdot (\l-1)^{\l-1}}
{(ps+\l-1)^{ps+\l-1}}\right\}(1+o(1))=
\frac{(\l+1)^{\l+1}}{2e^{\l-1}}\cdot
\frac{\log_2\,s}{s^{\l+1}}(1+o(1)),
$$
где учли, что
$$
\max_{0<p<1}\;
\{(1-p)^2\,p^{\l-1}\}\;=\;
(\l-1)^{\l-1}\,\frac{4}{(\l+1)^{\l+1}}
$$
и максимальное значение достигается при~$p=\frac{\l-1}{\l+1}$.

Теорема 3 доказана.$\quad \square$

\subsection{Выводы границ скорости СД $\;s_L$-кодов}
\subsubsection{Доказательство теоремы 6}
\quad
\textbf{Доказательство утверждения 1.} \quad
 Граница~\req{lb-triv1} означает, что при $s\leq L$ утверждение теоремы~6, т.е. неравенство
$R_L(s)\leq 1/s$,  верно. Пусть теперь $s> L\ge1$ и $X$ -- произвольный СД $\;s_L$-код объема $t$
и длины~$N$, который содержит хотя бы один  столбец (кодовое слово) произвольного
фиксированного веса $w$, $1\le w\le N$.
По аналогии с рассуждениями в~\cite{dr82}-\cite{dmtv02} для случая $L=1$, можем написать, что данный вес
\beq{EstW}
w\le N - N_{ld}(t-1, s-1, L),
\eeq
где $N_{ld}(t, s, L)$ обозначает минимальную длину СД $\;s_L$-кода объема~$t$ из определения~2.

Из~\req{EstW}, первого неравенства в~\req{lb-triv2}, а также
 верхних оценок \cite{dr82}-\cite{dmtv02} на количество слов фиксированного веса в
дизъюнктивном  $\len s/L\rin$-коде вытекает, что для объема~$t$ любого СД $\;s_L$-кода
справедлива верхняя граница:
\beq{EstT}
t\leq N_{ld}(t, s, L) + \sum\limits_{w=\len s/L\rin + 1}^{N_{ld}(t, s, L) - N_{ld}(t-1, s-1, L)}\frac{s^2}{L^2}\frac{{N_{ld}(t, s, L)\choose{\lceil q\rceil}}}{{\len q\rin\len s/L\rin\choose{\len q\rin}}} + L - 1,\quad q=\frac{w}{\len s/L\rin}.
\eeq

Будем здесь и далее использовать обозначения~\req{recL1}-\req{recL4} из формулировки теоремы~6
для описания  рекуррентной верхней границы~$\overline{R}_L(s)$.
Если $t\to\infty$, то с помощью~\req{EstT} и аналитических рассуждения, аналогичных~\cite{dr82} и \cite{dmtv02},
получаем
\beq{Est-t}
\frac{\log_2t}{N_{ld}(t, s, L)}
\leq\max_{0\leq v\leq 1-\frac{N_{ld}(t-1, s-1, L)}{N_{ld}(t, s, L)}}\,
f_{\len s/L\rin}(v)(1+o(1)), \quad s > L,\quad t\to\infty.
\eeq
Теперь покажем, что скорость СД $s_L$-кодов
$R_L(s)\leq \overline{R}_L(s)$ при $s>L$.
Для этого достаточно доказать от противного по индукции, что $R_L(s)\leq r_L(s)$ при $s>L$. Базой такой индукции будет неравенство
\beq{Base}
R_L(L+1)\leq r_L(L+1).
\eeq
Для доказательства \req{Base} достаточно показать, что при $s=L+1$ тривиальная граница будет лучше рекуррентной, то есть, что
\beq{trivBetter}
\frac{1}{L+1}\leq r_L(L + 1).
\eeq
Для каждого значения параметра $s=2,3,\dots$ введем вспомогательную функцию аргумента $x$, $0<x<1$:
\beq{Gsx}
G_s(x)\eq x-\max\limits_{0\leq v\leq1-\frac{x}{\overline{R}_L(s-1)}}f_{\len s/L\rin}(v),\quad 0<x<1.
\eeq
Непосредственно из определения \req{Gsx} следует, что  функция $G_s(x)$, $0<x<1$,
монотонно возрастает и ее единственным нулем является $r_L(s)$. Следовательно, для доказательства \req{trivBetter} достаточно показать, что $G_{L+1}\left(\frac{1}{L+1}\right)\leq 0$. Используя  неравенство \req{axin} получаем  оценку
$$
G_{L+1}\left(\frac{1}{L+1}\right)=\frac{1}{L+1}-
\max\limits_{0\leq v\leq1-\frac{L}{L+1}}f_1(v)\leq
\frac{1}{L+1}-f_1\left(\frac{1}{L+1}\right)\leq
$$
\beq{GUp}
\leq
\frac{1}{L+1}\(1-\log_2(L+1)+\frac{\log_2 e }{L+1}\)<0.
\eeq
Справедливость последнего неравенства при $L=2$ проверяется непосредственно подстановкой, а при больших $L$ оно верно из соображений монотонности. Таким образом, доказано неравенство \req{trivBetter}, а следовательно, и база индукции \req{Base}.

От противного проверим, что выполняется шаг индукции. Учитывая  определение~\req{RsL} скорости  СД $s_L$-кодов $R_L(s)$, напишем предположения индукции и противного, т.е.
\beq{Anti}
R_L(s-1)\,\eq\,\overline{\lim\limits_{t\to\infty}}\frac{\log_2t}{N_{ld}(t, s-1, L)}\leq \overline{R}_L(s-1),
\quad
R_L(s)\,\eq\,\overline{\lim\limits_{t\to\infty}}\frac{\log_2t}{N_{ld}(t, s, L)}> r_L(s).
\eeq
Тогда, в силу~\req{Anti}, верна цепочка неравенств
$$
\overline{\lim\limits_{t\to\infty}}\left(1-\frac{N_{ld}(t-1, s-1, L)}{N_{ld}(t, s, L)}\right)<
\overline{\lim\limits_{t\to\infty}}\left(1-\frac{N_{ld}(t-1, s-1, L)\cdot r_L(s)}{\log_2t}\right)\leq
1-\frac{r_L(s)}{\overline{R}_L(s-1)}.
$$
Из~\req{Est-t} и полученного неравенства следует, что скорость СД $s_L$-кодов
$$
R_L(s)=\overline{\lim\limits_{t\to\infty}}
\frac{\log_2t}{N_{ld}(t, s, L)}\leq
\overline{\lim\limits_{t\to\infty}}\max_{0\leq v\leq 1-\frac{N_{ld}(t-1, s-1, L)}{N_{ld}(t, s, L)}}
f_{\len s/L\rin}(v)
\le\max_{0\leq v\leq1-\frac{r_L(s)}{\overline{R}_L(s-1)}}
f_{\len s/L\rin}(v)=r_L(s).
$$
где в последнем равенстве применили определение~\req{recL3}-\req{recL4} величины~$r_L(s)$.
Полученное противоречие доказывает индукционный переход.

Теперь для завершения вывода утверждения 1 установим  равенство~\req{rLs}.
Будем рассуждать от противного. Так как  определяемая \req{entropy}-\req{rec1-1}
функция $f_{\len s/L\rin}(v)$, параметра $v$, $0<v<1$, выпукла вверх и достигает своего максимума при $v=v_{\len s/L\rin}$,
то предположение противного можно записать в виде:
\beq{rLsAnti}
r_L(s)=f_{\len s/L\rin}\(v_{\len s/L\rin}\), \quad v_{\len s/L\rin}<1-\frac{r_L(s)}{\overline{R}_L(s-1)}, \quad L\ge1,\quad s>2L.
\eeq
Отсюда следует
\beq{Wrong1}
v_{\len s/L\rin}< 1-\frac{f_{\len s/L\rin}\left(v_{\len s/L\rin}\right)}{\overline{R}_L(s-1)}
\quad\Longleftrightarrow\quad
\overline{R}_L(s-1)>\frac{f_{\len s/L\rin}\left(v_{\len s/L\rin}\right)}{1-v_{\len s/L\rin}}.
\eeq
Покажем, что \req{Wrong1} неверно, т.е. выполняется неравенство
\beq{RL1}
\overline{R}_L(s-1)\leq\frac{f_{\len s/L\rin}\left(v_{\len s/L\rin}\right)}{1-v_{\len s/L\rin}}, \quad L\ge1,\quad s>2L.
\eeq
Так как $v_{\len\frac{s-1}{L}\rin}$ является точкой, где функция $f_{\len\frac{s-1}{L}\rin}$ достигает своего глобального максимума, то верно следующее
$$
\overline{R}_L(s-1)\leq r_L(s-1)\leq f_{\len\frac{s-1}{L}\rin}\left(v_{\len\frac{s-1}{L}\rin}\right),\quad L\ge1,\quad s>2L.
$$
Поэтому  для вывода \req{RL1}  достаточно проверить, что
\beq{RL2}
f_{\len\frac{s-1}{L}\rin}\left(v_{\len\frac{s-1}{L}\rin}\right)\leq
\frac{f_{\len s/L\rin}\left(v_{\len s/L\rin}\right)}{1-v_{\len s/L\rin}},\quad L\ge1,\quad s>2L.
\eeq
Если $s\ne kL$, то $ \len\frac{s-1}{L}\rin = \len s/L\rin $, а потому справедливость~\req{RL2} очевидна.
Если  $s= kL$, то $ \len\frac{s-1}{L}\rin=k-1$, а $\len s/L\rin=k $. Тогда \req{RL2}
вытекает из неравенства 
$f_{k-1}(v_{k-1})\leq\frac{f_k(v_k)}{1-v_k}$, $ k > 2$, полученного  в~\cite{dr82}.


Утверждение 1 доказано полностью.

\textbf{Доказательство утверждения 2.} \quad
Будем использовать функцию $G_s(x)$, определенную формулой \req{Gsx}. С учетом
очевидного неравенства $\overline{R}_L(s-1)\le{1}/{(s-1)}$,  $s\ge2$, $L\ge1$,
свойства монотонного возрастания функции $f_{\len s/L\rin}(v)$, $0\le v\le1/s$,
вытекающего из оценки~\req{K2},
 и определений~\req{entropy}-\req{rec1},
получаем цепочку неравенств
\begin{multline*}
G_s\left(\frac{1}{s}\right)\geq\frac{1}{s}-\max_{0\leq v\leq1-\frac{s-1}{s}} f_{\len s/L\rin}(v)=
\frac{1}{s}-\max_{0\leq v\leq\frac{1}{s}} f_{\len s/L\rin}(v)=
\frac{1}{s}-f_{\len s/L\rin}\left(\frac{1}{s}\right)
>\\
>\frac{1}{s}-h\left(\frac{1}{s\len s/L\rin}\right)>\frac{1}{s}-\frac{1}{s\len s/L\rin}\log_2\left[s\len s/L\rin\right]-\frac{2\log_2e}{s\len s/L\rin}, \quad s>3.
\end{multline*}
Отсюда следует, что при фиксированном $L$, $L\ge1$, существует целое число $s(L)\ge3$, такое,
что последовательность
$G_s\left(\frac{1}{s}\right)>0$ при $s>s(L)$. Вкупе со свойством, отмеченным после~\req{Gsx}, это означает, что $r_L(s)<\frac{1}{s}$ при $s>s(L)$.
В частности, при достаточно больших $L$ и при $s>L\(\log_2L+3\log_2\[\log_2L\]\)$ будет выполнено неравенство
$$
\frac{1}{s}-\frac{1}{s\len s/L\rin}\log_2\left(e^2s\len s/L\rin\right)=
\frac{1}{s}\left(1-\frac{\log_2\len s/L\rin+\log_2\[e^2s\]}{\len s/L\rin}\right)>0.
$$
Таким образом, имеем $s(L)\le L\log_2L(1+o(1))$.

Для доказательства неравенства $s(L)\ge L\log_2L(1+o(1))$ проверим по индукции, что $G_s(1/s)< 0$ при $L<s<L \log_2L - L$.
 База индукции для $s=L+1$, т.е. неравенство \req{GUp}, была доказана при выводе утверждения~1.
Из предположения индукции, т.е. неравенства $G_{s-1}(1/(s-1))< 0$, вытекает, что $1/(s-1)< r_L(s-1)$ и,
следовательно,  верхняя граница скорости $\overline{R}_L(s-1)={1}/{(s-1)}$ совпадает с тривиальной.
Поэтому проверкой индукционного перехода служит  цепочка соотношений:
$$
G_s\left(\frac{1}{s}\right)=\frac{1}{s}-\max_{0\leq v\leq1-\frac{s-1}{s}}\,f_{\len s/L\rin}(v)\,=\,
\frac{1}{s}-f_{\len s/L\rin}\left(\frac{1}{s}\right)\le
\frac{1}{s}\(1-\frac{\log_2s-\frac{\log_2 e }{s\len s/L\rin}}{\len s/L\rin}\)<0,
$$
где в первом неравенстве для оценки последовательности $f_{\len s/L\rin}(1/s)$ воспользовались~\req{axin}, а во втором учли, что $L<s<L \log_2L - L$.

Таким образом, получили $s(L)= L\log_2L(1+o(1))$.


Утверждение 2 доказано.

\textbf{Доказательство утверждения 3.} \quad
Для фиксированного $L\ge1$ введем
последовательность $K_L(s)$, $s\ge L+1$, задаваемую рекуррентно:
\beq{KLs}
K_L(L)\eq1,\quad K_L(s)\eq\left[f_{\len s/L\rin}\left(1-\frac{K_L(s-1)}{K_L(s)}\right)\right]^{-1},\quad s=L+1,L+2,\dots.
\eeq
Из определений~\req{recL1}-\req{rLs} нетрудно увидеть, что при любом  фиксированном $L\ge1$
справедливы соотношения
$$
r_L(s)\leq\frac{1}{K_L(s)},\quad s\ge1,\quad \text{и}\quad
r_L(s)=\frac{1}{K_L(s)}(1+o(1))\quad \text{при}\quad s\to\infty.
$$

С помощью рассуждений, аналогичных выводу теоремы~1, можно установить неасимптотическую
 верхнюю границу
\beq{KLsUp}
K_L(s)\le\frac{(s+1)^2}{2L\log_2\[\frac{s+1}{8}\]},\quad L\ge1,\quad s\ge8.
\eeq
Следовательно, для доказательства  асимптотического равенства \req{RLas}  достаточно показать, что
имеет место асимптотическое неравенство
\beq{KLsDown}
K_L(s)\geq\frac{s^2}{2L\log_2s}(1+o(1)), \quad s\to\infty.
\eeq
При любом $s>L$ функция $f_{\len s/L\rin}(v)$  аргумента $v$, $0\leq v\leq 1$, выпукла вверх. Поэтому для любого
$a$, $0<a<1$, значение функции
\beq{ConvexUp}
f_{\len s/L\rin}(v)\leq f_{\len s/L\rin}(a)+(v-a)f'_{\len s/L\rin}(a), \quad 0\leq v\leq 1,\quad 0<a<1.
\eeq
Положим в ~\req{ConvexUp} число $v\eq1-\frac{K_L(s-1)}{K_L(s)}$. Затем, подставив правую часть
~\req{ConvexUp} в  определение~\req{KLs}, после элементарных преобразований приходим к неравенству
\beq{KLsConv}
K_L(s)\geq K_L(s-1)+\frac{1-g_{\len s/L\rin}(a)K_L(s-1)}{f'_{\len s/L\rin}(a)+g_{\len s/L\right]}(a)},\quad s>L,\quad 0<a<1.
\eeq
где
\beq{gDef}
g_{\len s/L\rin}(a)\eq f_{\len s/L\rin}(a)-af'_{\len s/L\rin}(a),\quad s>L,\quad 0<a<1.
\eeq
Для функций \req{rec1} и \req{gDef} в~\cite{dr82}-\cite{dmtv02}
были доказаны  свойства:
\beq{gUp}
g_{\len s/L\rin}\left(\frac{2}{\len s/L\rin}\right)\leq\frac{2\log_2e}{\len s/L\rin^2-2},\quad s>L,
\eeq
\beq{fgUp}
f'_{\len s/L\rin}\left(\frac{2}{\len s/L\rin}\right)+g_{\len s/L\rin}\left(\frac{2}{\len s/L\rin}\right)
\leq\frac{\log_2\left[\frac{\len s/L\rin}{2}\right]}{\len s/L\rin},\quad s>L.
\eeq
Также заметим, что для достаточно больших $s>s_0$ из \req{KLsUp} и \req{gUp}  следует неравенство
\beq{Positive}
1-g_{\len s/L\rin}\left(\frac{2}{\len s/L\rin}\right)K_L(s-1)>0.
\eeq
Если $s>s_0$, то полагая в \req{KLsConv} число $a\eq\frac{2}{\len s/L\rin}$ и
 принимая во внимание~\req{gUp}-\req{Positive},  получаем
\beq{KLsRec}
K_L(s)\geq K_L(s-1)+\frac{\len s/L\rin}{\log_2\left[\frac{\len s/L\rin}{2}\right]}-
K_L(s-1)\frac{2\len s/L\rin\log_2e}{\left(\len s/L\rin^2-2\right)\log_2\left[\frac{\len s/L\rin}{2}\right]}.
\eeq
Если $s\to\infty$,  то в силу \req{KLsUp}, для последнего слагаемого в правой части \req{KLsRec}
имеет место асимптотическая оценка:
$$
K_L(s-1)\frac{2\len s/L\rin\log_2e}{\left(\len s/L\rin^2-2\right)\log_2\left[\frac{\len s/L\rin}{2}\right]}
=o\(\frac{s}{\log_2s}\).
$$
Поэтому, при $s\to\infty$ рекуррентное неравенство \req{KLsRec} позволяет написать асимптотическую нижнюю
границу:
\beq{KLsDown1}
K_L(s)\geq\sum\limits_{k=2L}^s\frac{\len k/L\rin}{\log_2\left[\frac{\len k/L\rin}{2}\right]}(1+o(1))\geq
\sum\limits_{k=2L}^s\frac{k}{L\log_2s}(1+o(1))
=\frac{s^2}{2L\log_2s}(1+o(1)).
\eeq

Асимптотическое неравенство \req{KLsDown} является следствием \req{KLsDown1} .  Утверждение 3 доказано.

Теорема 6 доказана.$\quad\square$

\subsubsection{Доказательство теоремы 7}
\quad \textbf{Доказательство утверждения 1.} \quad Фиксируем параметр $Q$, $0<Q<1$, а также целочисленные параметры $s\ge1$ и
$L\ge1$, где $s+L<t$. Будем использовать обозначения, введенные в доказательстве теоремы~4.
Для любых непересекающихся множеств  $\S,\,\L\subset[t]$, $|\S|=\s$, $|\L|=L$, $\S\cap\L=\varnothing$,
 и кода $X=(\x(1),\x(2),\dots,\x(t))$, объема $t$ и длины $N$,
соответствующую пару $\left(\x(\S),\x(\L)\right)$ будем называть $(s_L)$-{\em плохой парой}, если
$$
\bigvee_{i\in\S}\x(i)\succeq \bigvee_{j\in\L}\x(j).
$$
 Для ансамбля   $E(N,t,Q)$ равновесных  кодов $X$ с весом $w\eq\lfloor QN\rfloor$
 символом $P_2(N,Q,s,L)$ обозначим  вероятность события:
<<пара $\left(\x(\S),\x(\L)\right)$ является $(s_L)$-плохой парой>>.
Определение~\req{RsL} скорости  $R_L(s)$ СД $s_L$-кодов и рассуждения,
аналогичные тем, которые были сделаны  в начале вывода теоремы~4, приводят к неравенству
\beq{First2}
R_L(s)\ge \underline{R}_L(s)\,\eq\,\frac{1}{s+L-1}\max_{0<Q<1}A_L(s,Q),
\quad
A_L(s,Q)\eq\overline{\lim_{N \to \infty}}\frac{-\log_2 P_2(N,Q,s,L)}{N}.
\eeq
Для завершения вывода утверждения 1 теоремы 7 остается вычислить в явном виде функцию $A_L(s,Q)$ и
показать, что она описывается соотношениями~\req{ranL2}-\req{ranL3}.

Используя терминологию типов (см. вывод теоремы~4), вероятность $P_2(N,Q,s,L)$
представим в виде:
\beq{PV}
P_2(N,Q,s,L)=\sum\limits_{\{n(\a)\}}\frac{N!}{\prod_{\a}n(\a)!}
{ {N-n(\0)} \choose \lfloor QN\rfloor}^L {N \choose \lfloor QN\rfloor}^{-s-L},
\eeq
где суммирование идет по всевозможным типам $\{n(\a)\}$, $\a\in\{0,1\}^s$, для которых
\beq{CondD}
0\le n(\a)\le N,\quad
\sum\limits_{\a}n(\a)=N,\quad
|\x(i)|=\sum\limits_{\a:\,a_i=1}n(\a)=\lfloor QN\rfloor \; \text{для любого }i\in[s].
\eeq
Пусть $N \to \infty$ и $n(\a)\eq N[\tau(\a)+o(1)]$, где
фиксированное распределение распределение вероятностей $\tau\eq\{\tau(\a)\}$, $\a\in\{0,1\}^s$,  обладает свойствами,
индуцированными условиями~\req{CondD}, т.е.
\beq{tau}
 0\le \tau(\a)\le1,\quad
 \sum\limits_{\a\in\{0,1\}^s} \tau(\a) = 1, \quad
 \sum\limits_{\a:\,a_i = 1} \tau(\a) = Q \quad \text{для любого } i\in[s].
\eeq
С помощью формулы Стирлинга для типов, соответствующих этому распределению, находим логарифмическую асимптотику
слагаемого в~\req{PV}:
$$
-\log_2\left\{\frac{N!}{\prod_{\a}n(\a)!}
{ {N-n(\0)} \choose \lfloor QN\rfloor}^L {N \choose \lfloor QN\rfloor}^{-s-L}\right\}\,=\,N[F(\tau,Q)+o(1)],
$$
где
\beq{FVl}
F(\tau,Q)\eq \sum\limits_{\a}\tau(\a)\log_2\[ \tau(\a)\]-
(1-\tau(\0))\,L\cdot h\left(\frac{Q}{1-\tau(\0)}\right)+(s+L)h(Q).
\eeq
Пусть при $\tau_Q=\{\tau_Q(\a)\}$ достигается минимум функции $F\eq F(\tau, Q)$ для данного $Q$. Тогда
главный член логарифмической асимптотики  суммы слагаемых~\req{PV} есть
\beq{Second2}
A_L(s,Q)\,\eq\,
\overline{\lim_{N \to \infty}}\frac{-\log_2 P_2(N,Q,s,L)}{N}=
\min_{\tau\in\req{tau}}\,F(\tau, Q)\,=\,F(\tau_Q,Q).
\eeq


Запишем  соответствующую задачу минимизации: $F\to \min$.
\begin{align}\label{Pro2}
&\text{Основная функция:}\quad\quad F(\tau,Q): \mathbb{Y}\to \mathbb{R}.& \notag\\
&\text{Ограничения:}\,\,
\begin{cases}
 \sum\limits_{\a\in\{0,1\}^s} \tau(\a) = 1;\\
 \sum\limits_{\a:\,a_i = 1} \tau(\a) = Q \quad \text{для любого } i\in[s].\\
 \end{cases}&\\
&\text{Область поиска } \mathbb{Y}:\quad
0<\tau(\a)<1,\quad \a\in\{0,1\}^s.\notag
\end{align}
Поиск точки минимума $\tau_Q$ будем вести стандартным методом множителей Лагранжа. Рассмотрим лагранжиан
$$
\Lambda \eq F(\tau, Q) + \lambda_0 \left(\sum_{\a} \tau(\a) - 1\right) + \sum_{i = 1}^s \lambda_i \left(\sum\limits_{\a:\,a_i = 1}\, \tau(\a) - Q\right).
$$
Тогда необходимые условия экстремального распределения имеют вид:
\begin{gather}\label{Cond3}
\begin{cases}
\frac{\partial\Lambda}{\partial(\tau(\a))} = \log_2\[ \tau(\a)\] + \log_2 e + \lambda_0 +
\sum_{i = 1}^s\, a_i \, \lambda_i=0 \quad \text{при } \a \neq\0;\\
\frac{\partial \Lambda}{\partial(\tau(\0))} = \log_2\[ \tau(\0)\] + \log_2 e + \lambda_0 +
L \log_2 \[\frac{1 - \tau(\0)}{1 - \tau(\0) - Q}\]=0.
\end{cases}
\end{gather}
Используя рассуждения между формулами~\req{Cond} и~\req{Cond2}
при доказательстве утверждения~1 теоремы~4, получаем:  если  существует решение
$\tau=\tau_Q=\{\tau_Q(\a)\}$ системы~\req{Cond3} при
ограничениях~\req{Pro2} в области $\mathbb{Y}$, то оно единственно и дает  минимум функции $F$ на $\mathbb{Y}$.
Кроме того,  экстремальное распределение $\tau_Q=\{\tau_Q(\x)\}$ удовлетворяет  условиям:
\begin{gather}\label{Cond4}
\begin{cases}
\mu + \nu \sum_{i = 1}^s \,a_i + \log_2 \tau(\a) = 0 \quad \text{при }\a \neq \0;\\
\mu + \log_2 \tau(\0) + L \log_2\[ \frac{1 - \tau(\0)}{1 - \tau(\0) - Q}\] = 0,
\end{cases}
\end{gather}
где
$$
\nu\eq\lambda_1=\lambda_2=\dots=\lambda_s,\quad\mu\eq\log_2 e + \lambda_0.
$$
После замены параметра $y\eq \frac{1}{1 + 2^{-\nu}}$, $0<y<1$, первое уравнения системы~\req{Cond4} дает
\beq{tauQ}
\tau(\a) =\frac{2^{-\nu \sum a_i}}{2^{\mu}}\,=\,
\frac{1}{2^{\mu}y^s}\,(1-y)^{\sum a_i}\,y^{s-\sum a_i}
\quad\text{при}
\quad \a \neq \0.
\eeq
Ограничение  в задаче~\req{Pro2} на распределение вероятностей~\req{tauQ}  приводит к равенству
$$
Q = \sum\limits_{\a:\,a_i = 1} \tau_Q(\a)=\frac{1}{2^{\mu}y^s}\,\sum\limits_{\a:\,a_i = 1} \,(1-y)^{\sum a_j}\,y^{s-\sum a_j}=
$$
$$
=\frac{1}{2^{\mu}y^s} \sum_{k = 0}^{s-1} {s-1 \choose k} y^{s - k - 1}(1-y)^{k + 1} =
\frac{1 - y}{2^{\mu}y^s},\quad\text{для любого } i\in[s].
$$
 При  фиксированном  $Q$, $0<Q<1$, данное равенство представляет собой уравнение
 связи между параметрами $\mu$ и~$y$, описывающими экстремальное распределение~$ \tau_Q=\{\tau_Q(\a)\} $:
\beq{ay}
\frac{1}{2^{\mu}y^s}=\frac{Q}{1-y}
\qquad\Leftrightarrow\qquad \mu = \log_2 \[\frac{1 - y}{Qy^s}\].
\eeq
Применяя~\req{ay}, для распределения вероятностей~\req{tauQ} вычислим
$$
1 - \tau(\0) = \sum_{\a \neq \0} \tau(\a) =
\frac{1}{2^{\mu}y^s} \sum_{k = 1}^s {s \choose k}y^{s - k} (1 - y)^k = \frac{Q(1-y^s)}{1 - y}.
$$
Таким образом, после исключения параметра $\mu$ компоненты
экстремального распределения~\req{tauQ} принимают вид функций одной и той
независимой переменной  $y$, $0<y<1$:
\beq{tauQ1}
\tau_Q(\a) = \frac{Q}{1 - y}\, y^{[s - \sum a_i]} (1 - y)^{\sum a_i}\quad\text{при } \a\neq0; \quad \tau_Q(\0) =1- \frac{Q(1 - y^s)}{1 - y}.
\eeq

Подставляя~\req{tauQ1} во второе уравнение системы~\req{Cond4} и учитывая~\req{ay},
приходим к равенству
\beq{ranL4}
\log_2\[\frac{1 - y - Q(1 - y^s)}{Qy^s}\] + L \log_2\[ \frac{1 - y^s}{y - y^s}\] = 0.
\eeq
которое равносильно уравнению~\req{ranL3} для  параметра $y$, $1-Q<y<1$,
из формулировки утверждения~1 теоремы~7. Уравнение~\req{ranL4} имеет
единственное решение $y=y(s,Q)$, поскольку для рассматриваемой нами выпуклой задачи Лагранжа существует
единственное экстремальное распределение~\req{tauQ1}, задаваемое параметром $y$,~$0<y<1$.

Чтобы найти значение $F(\tau_Q, Q)$ для
искомого минимума в~\req{Second2}, подставим вероятности~\req{tauQ1} в определение~\req{FVl}
функции~$ F(\tau, Q)$. Затем группируя слагаемые в~\req{FVl} по $s$ и  $L$,
вычисляем  $F(\tau_Q, Q)$  как функцию независимой переменной $y$,~$0<y<1$. Используя символ расстояния Кульбака~\req{Kul},
результат можно записать в виде:
\beq{LastV}
F(\tau_Q, Q) =
\log_2\[ \frac{Q}{1-y}\] - sK\left( Q, 1-y \right) - LK\left( Q, \frac{1-y}{1-y^s} \right).
\eeq
Из~\req{First2},~\req{Second2},~\req{ranL4} и~\req{LastV} вытекает приведенная в утверждении~1 теоремы~7
нижняя оценка~\req{ranL1}-\req{ranL3} для  скорости СД $s_L$-кодов.

Утверждение 1 доказано.

\textbf{Доказательство утверждения 2.} \quad При фиксированных $s\ge2$ и $L\ge1$
уравнение~\req{ranL3} будем интерпретировать как функцию $Q= Q_L(y,s)$   аргумента $y$, $0<y<1$, т.е.
\beq{ranL5}
Q = Q_L(y,s)\,\eq\,\frac{1 - y}{1 - r_L(y,s)}, \quad
r_L(y,s)\,\eq\,  y^s
\left[ 1 - \left( \frac{y - y^s}{1 - y^s} \right)^L \right],\qquad 0<y<1.
\eeq
Тогда, применяя явную формулу~\req{Kul} для расстояния Кульбака,
определение границы случайного кодирования~\req{ranL1}-\req{ranL3}
 можно переписать в виде
\beq{ranL6}
\underline{R}_L(s)\,\eq\,\frac{1}{s+L-1}\max\limits_{0<y<1}\,T_L(y,s),
\eeq
где
\begin{multline}\label{ranL7}
T_L(y,s)\,\eq\,(1 - sQ-LQ) \log_2 \[\frac{Q}{1 - y}\] - 
\\
- (s+L)(1-Q) \log_2\[ \frac{1 - Q}{y}\]
-L\log_2\[1 - y^s\] +L(1-Q) \log_2\[ 1 - y^{s-1}\]
\end{multline}
и параметр  $Q$ в правой части~\req{ranL7} определяется~\req{ranL5}.

Пусть $L\ge1$ фиксировано и  $s\to\infty$. Если в определениях~\req{ranL5} и~\req{ranL7} положить
 $y=1-c/s$, где параметр $c=c_L>0$ не зависит от $s$,  то~\req{ranL6} означает, что граница
случайного кодирования
\beq{1ranL}
\underline{R}_L(s)\,\ge\,\frac{1}{s+L-1}\,T_L\left(1-\frac{c}{s},s\right),\quad c<s.
\eeq
Вычисление главных членов асимптотических разложений в
формулах~\req{ranL5} и~\req{ranL7}, когда  $y=1-c/s$  и  $s\to\infty$, приводят к асимптотическим равенствам
\beq{2ranL}
Q = Q_L\left(1-\frac{c}{s},s\right)=\frac{c}{s}\,(1+o(1)),\quad
T_L\left(1-\frac{c}{s},s\right)=-\frac{L}{s}\,c \cdot \log_2\,(1-e^{-c})\,(1+o(1)).
\eeq
Несложно проверить, что при $c=\ln2=\frac{1}{\log_2e}=0,619$ достигается
\beq{3ranL}
\max\limits_{c\,>\,0}\,\{-c \cdot \log_2\,(1-e^{-c})\}=\frac{1}{\log_2e}.
\eeq
Поэтому из~\req{1ranL}-\req{3ranL} для границы случайного кодирования~\req{ranL5}-\req{ranL7} вытекает
асимптотическое неравенство
\beq{4ranL}
\underline{R}_L(s)\,\ge\,
\frac{L}{s^2\cdot\log_2e}\,(1+o(1)),\quad s\to\infty,\; L=1,2,\dots.
\eeq
 Подстановка значения $y=1-\ln2/s$  в~\req{ranL5} и вычисление соответствующего
значения параметра $Q$  приводит к асимптотической формуле~\req{QLs}
для доли веса $Q_L(s)$ кодовых слов в равновесном ансамбле кодов, на котором достигается
асимптотика скорости, описываемая  правой частью~\req{4ranL}.

Для доказательства равенства~\req{ranL-4} и
завершения вывода  утверждения~2 надо, пользуясь определениями~\req{ranL5}-\req{ranL7},
проверить справедливость асимптотического равенства в~\req{4ranL}. Из данного равенства будет следовать,
что с помощью равновесного ансамбля кодов построенную в утверждении~1 нижнюю границу $\underline{R}_L(s)$
для  скорости  СД $s_L$-кодов нельзя существенно улучшить. Эта проверка здесь не приводится
по причине громоздкости используемых  рассуждений, а также  не очень большой значимости результата,
 который она дает.

Утверждение~2 доказано.

\textbf{Доказательство утверждения 3.} При фиксированных $ s \ge 2 $, $ L \ge 1 $ и параметре  $ c > 0 $,
не зависящем от $L$, рассмотрим уравнение
\beq{yFromEdge}
\( \frac{y - y^s}{1 - y^s} \)^L = c(1 - y), \quad 0 < y < 1.
\eeq
В силу свойств возрастания левой части и убывания правой части по $y$,  уравнение~\req{yFromEdge}  имеет
ровно один корень на интервале $ y \in (0, 1) $. Обозначим этот корень через~$y_L(s, c)$.
Подставив его в~\req{ranL5},  введем величины $Q=Q_L(s, c)$ и $r=r_L(s, c)$,
которые вместе с $y = y_L(s, c)$  будем интерпретировать как последовательности аргумента~$L=1,2,\dots$.

Пусть $ s \ge 2 $ фиксировано и $ L \to \infty $. Очевидны следующие
асимптотические свойства данных последовательностей:
\begin{align}
&y=y_L(s, c) = 1 + o(1), \nonumber \\
&r_L(s, c) = 1 - (s+c)(1-y)+o(1-y), \label{QryFromEdge}\\
&Q_L(s, c) = \frac{1}{s + c} (1 + o(1))  , \quad L \to \infty, \quad s = 2, 3, ..., \quad c> 0. \nonumber
\end{align}

Определение~\req{ranL5}-\req{ranL7}
означает, что для любого $c>0$ граница случайного кодирования
\beq{2ranL}
\underline{R}_L(s)\,\ge\,\frac{1}{s+L-1}\,T_L\left(y_L(s,
c),s\right),\quad c > 0. \eeq где величина $T_L\left(y_L(s,
c),s\right)$ задается  формулой~\req{ranL7}, в которой $y = y_L(s,
c)$, а $Q=Q_L(s, c)$. Применяя определение~\req{yFromEdge} и
свойства~\req{QryFromEdge}, можно вычислить главный член асимптотики
в правой части \req{ranL7} при $y = y_L(s, c)$ и $Q=Q_L(s, c)$, а
затем получить асимптотическое равенство:
\begin{multline}\label{FSubQ}
\frac{T_L\left(y_L(s, c),s\right)}{L} = \bigg( \log_2\[ \frac{s + c}{s}\]-\frac{s+c-1}{s + c} \log_2 \[\frac{s + c-1}{s - 1}\] -
\\
-
\frac{c}{s + c} \log_2\[ \frac{s - 1}{s}\] \bigg) (1+o(1)), \quad L\to\infty.
\end{multline}

С помощью взятия производной по $c$ нетрудно убедиться, что максимум правой части равенства~\req{FSubQ}
достигается при значении $c=c(s)\eq\frac{s^s - (s-1)^s}{(s-1)^{s-1}}$. Если теперь данное $c=c(s)$ подставить
 в~\req{FSubQ},  то пользуясь неравенством~\req{2ranL}, для границы случайного кодирования~\req{ranL5}-\req{ranL7}
устанавливаем  неравенство:
\beq{EdgeSub}
\underline{R}_{\infty}(s)\eq\lim\limits_{L\to\infty}\,
\underline{R}_{L}(s) \ge \log_2 \[ \frac{(s-1)^{s-1}}{s^s} + 1 \],
 \quad s = 2, 3,\dots.
\eeq
Если же $c=c(s)$ подставить в  формулу  для $Q$ из~\req{QryFromEdge}, то придем к асимптотической
формуле~\req{QsL} для  доли веса $Q_L(s)$ кодовых слов в равновесном ансамбле, на котором достигается
асимптотика скорости, описываемая  правой частью~\req{EdgeSub}.

Для доказательства равенства~\req{ranL-5} и завершения вывода  утверждения~3 надо, пользуясь~\req{ranL5}-\req{ranL7},
проверить справедливость  равенства в~\req{EdgeSub}. Эта проверка в данной работе не приводится
по той же причине, что была указана выше для утверждения~2.

Утверждение~3 доказано.

Теорема~7 доказана.$\quad \square$

\end{document}